\documentclass[aps,amsfonts,amsmath,prd,preprint,nofootinbib]{revtex4-1}
\usepackage{epsfig,bbm,cancel}
\usepackage{slashed,xcolor}
\usepackage[normalem]{ulem} % to use strikethrough 
\usepackage{amssymb}

\begin{document}

\title{BPS Skyrme Submodels of The Five-dimensional Skyrme Model}

\author{Emir Syahreza Fadhilla$^1$}
\email{emirsyahreza@students.itb.ac.id}
\author{Bobby Eka Gunara$^1$}
\email{bobby@fi.itb.ac.id (Corresponding author)}
\author{Ardian Nata Atmaja$^2$}
\email{ardi002@lipi.go.id}

\affiliation{$^1$Theoretical High Energy Physics  Research Division, Institut Teknologi Bandung,
Jl. Ganesha 10 Bandung 40132, Indonesia.}
\affiliation{$^2$ Research Center for Physics, Indonesian Institute of Sciences (LIPI),
Kompleks PUSPIPTEK Serpong, Tangerang 15310, Indonesia.}

\begin{abstract}
 In this paper, we search for the BPS skyrmions in some BPS submodels of the generalized Skyrme model in five-dimensional spacetime using the BPS Lagrangian method. We focus on the static solutions of the Bogomolny's equations and their corresponding energies with topological charge $B>0$ is an integer. We consider two main cases based on the symmetry of the effective Lagrangian of the BPS submodels, i.e. the spherically symmetric and non-spherically symmetric cases. For the spherically symmetric case, we find two BPS submodels. The first BPS submodels consist of a potential term and a term proportional to the square of the topological current. The second BPS submodels consist of only the Skyrme term. The second BPS submodel has BPS skyrmions with the same topological charge $B>1$, but with different energies, that we shall call ``topological degenerate'' BPS skyrmions. It also has the usual BPS skyrmions with equal energies, if the topological charge is a prime number. Another interesting feature of the BPS skyrmions, with $B>1$, in this BPS submodel, is that these BPS skyrmions have non-zero pressures in the angular direction. For the non-spherically symmetric case, there is only one BPS submodel, which is similar to the first BPS submodel in the spherically symmetric case. We find that the BPS skyrmions depend on a constant $k$ and for a particular value of $k$ we obtain the BPS skyrmions of the first BPS submodel in the spherically symmetric case. The total static energy and the topological charge of these BPS skyrmions also depend on this constant. We also show that all the results found in this paper satisfy the full field equations of motions of the corresponding BPS submodels.
\end{abstract}

\maketitle
\thispagestyle{empty}
\setcounter{page}{1}
\tableofcontents

\section{Introduction}
\label{sec:intro}

The Skyrme model was initially designed to be a model of the nucleon~\cite{Skyrme:1961vq,Skyrme:1962vh}. It has topological soliton solutions, known as skyrmions, which can be interpreted as baryons. As topological solitons, the skyrmions are characterized by a conserved topological charge $B$ which is identified as the baryon number~\cite{Witten:1983tx,Dothan:1986ae}. The energies of these skyrmions satisfy the Faddeev-Bogomolny lower energy bound $E\geq |B|$~\cite{Faddeev:1976pg}. Unfortunately, this bound can not be saturated by skyrmions with $B\neq0$~\cite{manton1987}. Under a particular spherically symmetric ansatz, it has been shown that a unit skyrmion (the nucleon), with $B=1$, has an energy about 23\% higher than its Faddeev-Bogomolny lower bound~\cite{Adkins:1983ya}. For other skyrmions with $B>1$, their binding energies are even higher than the multiple unit skyrmions and hence mostly unstable~\cite{Manton:2004tk}. Nevertheless, skyrmions are not only important in high energy and nuclear physics but also play an important role in other areas of physics from solid-state physics to nonlinear optics~\cite{Donati14926,Khawaja:2001zz,Baskaran:2011wg,Kiselev_2011,Fukuda_2011}.

Several attempts have been made to find (more) stable skyrmions, in particular, with $B>1$. These may include imposing non-spherically symmetric ansatzs~\cite{Braaten:1989rg,Hitchin:1995qw,Houghton:1995bs}. Other approaches are by considering additional higher-order terms other than the Skyrme term. A good mathematical description of skyrmion, found in \cite{manton1987}, describes the skyrmions as a topologically nontrivial map from a Riemannian manifold to another. This mathematical description becomes the key to generalize the Skyrme model by adding the higher-order term, first done in \cite{Marleau:1989fh}, and it has been turned recently into a combinatoric problem in \cite{Gudnason_2017}. A simple extension to the Skyrme model is by adding a term proportional to the square of the topological current density~\cite{Adkins:1984cf}. This additional term turns out to be responsible for stabilizing the skyrmions~\cite{Jackson:1985yz}. Furthermore, BPS skyrmions, which saturate the Faddeev-Bogomolny bound, are found in a model consist only of this term and a potential term~\cite{Adam:2010fg}. Although initially the construction of higher-order term is given for four-dimensional spacetime, a possible generalization in higher dimensional spacetime is demonstrated in \cite{Brihaye:2017wqa}. In that paper, they only studied unit skyrmions (static and stationary), with $B=1$, since the main goal was to find and study rotating black holes with Skyrme hair in five-dimensional spacetime.

The main purpose of this work is to find other possible BPS skyrmions in the generalized five-dimensional Skyrme model considered in~\cite{Brihaye:2017wqa}. In particular, we will try to find BPS skyrmions, for any integer $B$, in some (genuine) submodels of the generalized five-dimensional Skyrme model via the BPS Lagrangian method, which was initially used to find BPS vortices in three-dimensional spacetime models~\cite{Atmaja:2015umo}. The method not only succeeded in re-deriving the known BPS vortices, but it was also able to find new Bogomolny equations for other BPS vortices~\cite{Atmaja:2018ddi}. It has also been used to find the BPS monopoles and dyons in various $SU(2)$ Yang-Mills-Higgs models in the four-dimensional spacetime~\cite{Atmaja:2018cod,Atmaja:2020iti}. The reason for using this method is that it has been successfully used to find the BPS skyrmions in some (genuine and nongenuine) submodels of the generalized Skyrme model in the four-dimensional spacetime~\cite{Atmaja:2019gce}\footnote{We call genuine submodels of a generalized Skyrme model for submodels that can be obtained by turning off one or more couplings in the generalized Skyrme model, while non-genuine ones that cannot.}. Two main cases considered here are the spherically symmetric and non-spherically symmetric cases, which are defined to be the symmetry of the static energy density in a particular ansatz.

In section \ref{sec:model} we describe briefly the Skyrme model and the ansatz that we use in this paper. Then, in section \ref{sec:BPS1} we discuss submodels in the spherically symmetric Lagrangian density which is divided into two subsections, submodel with $\lambda_0=\lambda_1=\lambda_3=\lambda_4=0$ and submodel with $\lambda_1=\lambda_2=\lambda_3=0$. Submodels in the non-spherically symmetric Lagrangian are discussed in section \ref{sec:BPS2} which is divided into two subsections, submodel with one derivative term and submodels with four derivative terms. We conclude the results in section \ref{sec:concl}. Finally,  we put some additional computations that support our results in the Appendix.
%%%%%%%
%%%%%NEW SECTION

\section{Static $O(5)$ Skyrme Model}\label{sec:model}
As we know, the standard static Skyrme Model in \(3+1\) dimension is a \(\mathbb{R}^3\rightarrow S^3\) mapping
with the degree equals to one, in which the dynamical equation minimizes the following action functional
\begin{equation}\label{action4}
    \mathcal{S}_4=-\int d^4x \sqrt{-g}\left[\gamma_1 g^{ij}~Tr(R_i R_j)+\gamma_2g^{ij}g^{kl}~Tr([R_i, R_k][R_j, R_l])\right] ~ ,
\end{equation}
with \(R_i \equiv (\partial_i U) U^{\dagger} \) where \(U=\phi^0I_{2}+i\vec{\phi}.\vec{\sigma}\) is an \(SU(2)\) valued chiral field \cite{Manton:2004tk, Brihaye:2017wqa} and $i,j,k,l$ are spacetime indices. \(\vec{\sigma}=(\sigma_1,\sigma_2,\sigma_3)\) are Pauli matrices and \(\phi=(\phi^0,\vec{\phi})\) are real scalar fields that  satisfy the \(O(4)\) model condition, \(\phi^a \phi^a =1\). This is the original model proposed by Skyrme \cite{Skyrme:1961vq, Skyrme:1962vh}. On the other hand, one can construct such a functional by introducing strain tensor \(D=JJ^T\), where \(J\) is the Jacobean matrix of the map \cite{manton1987}. This strain tensor characterizes the distortion induced by the map. The energy functional of the static Skyrme model can be constructed from the invariant of the strain tensor, which are combinations of its eigenvalues. As a result of these arguments we can rewrite the Lagrangian in \eqref{action4} as a sum of these two terms, namely
\begin{eqnarray}
    \phi^a_i\phi^a_jg^{ij} ~ ,\\
    \phi^a_{[i}\phi^b_{j]}\phi^a_{[k}\phi^b_{l]}g^{ij}g^{kl} ~ ,
\end{eqnarray}
where \(\phi^a_i\equiv\frac{\partial\phi^a}{\partial x^i}\) (with $a,b = 0,...,3$) and \(g^{ij}\) are components of the inverse of the spacetime metric. This construction of the Skyrme model gives rise to the generalization for the higher order terms given in \cite{Gudnason_2017} and the Skyrme model in five dimensions given in \cite{Brihaye:2017wqa}.

Suppose we have \(D\) eigenvalues from a \(d+1\) dimensional Skyrme Model as \(\lambda_1^2,\lambda_2^2,\dots,\lambda_d^2\). The corresponding Lagrangian can be written in the most general way as
\begin{equation}
    L=\gamma_0 V+\sum_{n=1}^d \gamma_nL_n ~ ,
\end{equation}
where \(\gamma_0, \gamma_n \in\mathbb{R}\) are coupling constants, \(V\equiv V(\phi)\) is an arbitrary potential, and \(L_n\) satisfy
\begin{eqnarray}
    L_1&\propto& \lambda_1^2+\dots+\lambda_d^2 ~ ,\\
    L_2&\propto& \lambda_1^2\lambda_2^2+\lambda_1^2\lambda_3^2+\dots+\lambda_{d-1}^2\lambda_d^2 ~ ,\\
    \vdots\nonumber\\
    L_d&\propto& \lambda_1^2\lambda_2^2\dots\lambda_d^2 ~ .
\end{eqnarray}
In four-dimensional case (\(d=3\)), this gives the generalized Skyrme model which is just the traditional model of Skyrme added with the BPS-Skyrme term (\(L_3\propto \lambda_1^2\lambda_2^2\lambda_3^2\)) proposed in \cite{Adam:2010zz}. By following the prescription given in the four-dimensional model, we can construct \(L\) for arbitrary spacetime dimensions with 
\begin{eqnarray}
    L_n&=&\frac{F^{2n}}{\left(n!\right)^2} ~ ,\\
    F^{2n}&\equiv&\phi^{a_1}_{[i_1}\dots\phi^{a_n}_{i_n]}\phi^{a_1}_{[j_1}\dots\phi^{a_n}_{j_n]}g^{i_1j_1}\dots g^{i_nj_n} ~ ,
\end{eqnarray}
where $a_1, ...,a_n = 1,...,d+1$ are internal indices, whereas $i$s and $j$s are spacetime indices, and $\phi^{a_1}_{[i_1}\dots\phi^{a_n}_{i_n]} $ is totally antisymetric tensor \cite{Brihaye:2017wqa}. Therefore the Lagrangian of the model can be written in the definition above as
\begin{equation}\label{actiond+1}
    L=\gamma_0 V+\sum_{n=1}^d \frac{\gamma_n}{\left(n!\right)^2}\phi^{a_1}_{[i_1}\dots\phi^{a_n}_{i_n]}\phi^{a_1}_{[j_1}\dots\phi^{a_n}_{j_n]}g^{i_1j_1}\dots g^{i_nj_n}.
\end{equation}
This is just an \(O(d+1)\) model generalization as proposed in \cite{Brihaye:2017wqa} if we take the potential to be \(V=1-\phi^{d+1}\). As a summary, the action functional for \(d+1\) dimensional Skyrme model takes the form
\begin{equation}
    \mathcal{S}_{d+1}=-\int dx^{d+1}\sqrt{-g}\left[\gamma_0 V+\sum_{n=1}^d \frac{\gamma_n}{\left(n!\right)^2}\phi^{a_1}_{[i_1}\dots\phi^{a_n}_{i_n]}\phi^{a_1}_{[j_1}\dots\phi^{a_n}_{j_n]}g^{i_1j_1}\dots g^{i_nj_n}\right].
\end{equation}

Now let us focus on  the five-dimensional Skyrme model discussed in \cite{Brihaye:2017wqa} which is the main discussion of the paper. In this section we review shortly about it. In particular, we consider a static case in which the spacetime metric can be written down in terms of bipolar spherical coordinates 
\begin{equation}\label{eq:metricanz}
ds^{2}=-dt^{2}+dr^{2}+r^{2} \left( d \theta ^{2}+\sin ^{2} \theta~ d\varphi_{1}^{2}+\cos ^{2} \theta~ d\varphi_{2}^{2} \right) ~ ,
\end{equation}
where $-\infty<t<\infty$ and $0\leq r<\infty$ are the time and radial coordinates, respectively, while $0\leq\theta\leq{\pi\over 2}$ is the polar coordinate and $0\leq\varphi_{1,2} < 2\pi$  are the azimuthal coordinates. Then, the form of the effective Lagrangian density and the energy-momentum tensor will be cast in terms of these coordinates using \eqref{eq:metricanz}. 

The five-dimensional Skyrme model can be thought of as an $O(5)$ sigma model of real fields $\phi^a$, $a=1,\ldots,5$, satisfying the constraint  \(  \phi ^{a} \phi ^{a}=1 \) \cite{Brihaye:2017wqa}. Here, we consider the generalized Skyrme model with the following Lagrangian density
\begin{equation} \label{eq:L} 
L_{eff} = \lambda _{0}V \left(  \phi  \right) + \lambda _{1}F^{2}+\frac{ \lambda _{2}}{4}F^{4}+\frac{ \lambda _{3}}{36}F^{6}+\frac{ \lambda _{4}}{576}F^{8} ~ ,
\end{equation} 
where $\lambda_n$ is the coupling of the $2n$-th order term, with $n=0,\ldots,4$. Here $V$ is the potential (zeroth order) term.
We choose a possible static ansatz for the scalar field that satisfies the constraint \(  \phi ^{a} \phi ^{a}=1 \), that is 
 \begin{subequations}\label{ansatz}
  \begin{align}
   \phi ^{1}+i \phi ^{2}&=\sin  \xi  \frac{f  }{\sqrt[]{1+f^{2}  }}e^{ig_{1}  } ~ ,\\
  \phi ^{3}+i \phi ^{4}&=\sin  \xi  \frac{1}{\sqrt[]{1+f^{2}}}~e^{ig_{2} } ~ ,\\
  \phi ^{5}&=\cos  \xi ~ ,
  \end{align}
 \end{subequations}
 where $\xi  \equiv \xi (r)$, $f \equiv f (\theta)$, $ g_1 \equiv g_1 \left(  \varphi _{1} \right) $, and $ g_2 \equiv g_2 \left(  \varphi _2 \right) $ are defined to be the effective fields such that the Lagrangian density \eqref{eq:L} can be recast into an effective Lagrangian density
\begin{equation}
L_{eff}= L_{0}+L_{1}+L_{2}+L_{3}+L_{4} ~ ,\label{L_eff}
 \end{equation}
where 
\begin{equation}
 L_{0}=\lambda _{0}V \left(  \phi  \right) ~,  
\end{equation}
\begin{equation} 
\begin{aligned} 
L_{1}=\lambda _{1} \left[  \left(  \xi'  \right) ^{2}+\frac{\sin ^{2} \xi }{r^{2}} \left( \frac{f' }{1+f^{2} } \right) ^{2}+\frac{\sin ^{2} \xi  }{r^{2}\sin ^{2} \theta ~}\frac{ \left( f  g_{1}' \right) ^{2}}{1+f^{2} }+\frac{\sin ^{2} \xi  }{r^{2}\cos ^{2} \theta ~}\frac{ \left( ~g_{2}' ~ \right) ^{2}}{1+f^{2}  } \right] ~, 
\end{aligned}\end{equation}
\begin{eqnarray}
 L_{2}&=&\lambda _{2} \left[ \frac{\sin ^{2} \xi }{r^{2}}\frac{1}{ \left( 1+f^{2} \right) ^{2}} \left(  \xi'f'  \right) ^{2}+\frac{\sin ^{2} \xi }{r^{2}\sin ^{2} \theta ~}\frac{ f^2 }{ \left( 1+f^{2} \right) } \left(  \xi'g_{1}' \right) ^{2}+\frac{\sin ^{2} \xi }{r^{2}\cos ^{2} \theta ~}\frac{1}{ \left( 1+f^{2}\right) } \left(  \xi'g_{2}'\right) ^{2}\right.\nonumber \\ 
&&\qquad\left.+\frac{\sin ^{4} \xi }{r^{4}\sin ^{2} \theta ~}\frac{  f^2 }{ \left( 1+f^{2} \right) ^{3}} \left( f'g_{1}' \right) ^{2}+\frac{\sin ^{4} \xi }{r^{4}\cos ^{2} \theta ~}\frac{1}{ \left( 1+f^{2} \right) ^{3}} \left( f'g_{2}'\right) ^{2}\right.\nonumber\\
&&\qquad\left.+\frac{\sin ^{4} \xi}{r^{4}\sin ^{2} \theta \cos ^{2} \theta }\frac{ f^2 }{ \left( 1+f^{2} \right) ^{2}} \left( g_{1}'g_{2}' \right) ^{2} \right],
\end{eqnarray}
\begin{eqnarray}
 L_{3}&=&\lambda _{3} \left[ \frac{1}{r^{4}\sin ^{2} \theta }\frac{  f^2  }{ \left( 1+f^{2} \right) ^{3}}\sin ^{4} \xi \left(  \xi'f'g_{1}' \right) ^{2}+\frac{1}{r^{4}\cos ^{2} \theta }\frac{1}{ \left( 1+f^{2}  \right) ^{3}}\sin ^{4} \xi \left(  \xi'f'g_{2}' \right) ^{2}\right.\nonumber \\ 
&&\qquad\left.+\frac{1}{r^{6}\sin ^{2} \theta \cos ^{2} \theta }\frac{ f^2 }{ \left( 1+f^{2} \right) ^{4}}\sin ^{6} \xi  \left(f'g_{1}'g_{2}'\right) ^{2}\right.\nonumber\\
&&\qquad\left.+\frac{1}{r^{4}\sin ^{2} \theta \cos ^{2} \theta }\frac{ f^2 }{ \left( 1+ f^{2}\right) ^{2}}\sin ^{4} \xi  \left(  \xi'g_{1}'g_{2}' \right) ^{2} \right] ~ ,
\end{eqnarray}
\begin{equation} L_{4}=\lambda _{4} \left[ \frac{1}{r^{6}\sin ^{2} \theta \cos ^{2} \theta }\frac{f^{2} }{ \left( 1+f^{2} \right) ^{4}}\sin ^{6} \xi \left(  \xi'f' g_{1}' g_{2}'  \right) ^{2} \right] ~ . \end{equation}
 where $\xi'  \equiv \frac{d\xi}{dr}$, $f' \equiv \frac{df}{d\theta} $, $ g_1' \equiv \frac{dg_1}{d\varphi_1}$, and $ g_2' \equiv \frac{dg_2}{d\varphi_2}$ are derivatives of the effective fields with respect to their arguments.
Using the ansatz (\ref{ansatz}), the topological charge has a simple form of
\begin{equation}
 B=-\frac{3}{8 \pi^{2}} \int \sin ^{3} \xi \frac{f}{\left(1+f^{2}\right)^{2}} d \xi d f d g_{1} d g_{2} ~ ,\label{top charge}
\end{equation}
We would also like to point out the expression for the components of the energy-momentum tensor for this model as follows 
\begin{equation}\label{energymomentum}
    T_{ij}=\lambda_0 T_{ij}^{(0)}+\lambda_1 T_{ij}^{(1)}+\lambda_2 T_{ij}^{(2)}+\lambda_3 T_{ij}^{(3)}+\lambda_4 T_{ij}^{(4)} ~ ,
\end{equation}
with
\begin{eqnarray}
  T_{ij}^{(0)}&=&-g_{ij}V ~ ,\\
  T_{ij}^{(1)}&=&2\phi _{i}^{a} \phi _{j}^{a}-g_{ij}F^2 ~ ,\\
  T_{ij}^{(2)}&=&\phi _{ik_1}^{ab} \phi _{jk_2}^{ab}g^{k_1k_2}-\frac{1}{4}g_{ij}F^4 ~ ,\label{SEMT L2}\\
  T_{ij}^{(3)}&=&\frac{1}{6}\phi _{ik_1l_1}^{abc} \phi _{jk_2l_2}^{abc}g^{k_1k_2}g^{l_1l_2}-\frac{1}{36}g_{ij}F^6 ~ ,\\
  T_{ij}^{(4)}&=&\frac{1}{72}\phi _{ik_1l_1m_1}^{abcd} \phi _{jk_2l_2m_2}^{abcd}g^{k_1k_2}g^{l_1l_2}g^{m_1m_2}-\frac{1}{576}g_{ij}F^8 ~ ,
\end{eqnarray}
that will be useful for our analysis in this paper. It is straightforward to show that \(T_{ij}=0\) for \(i\neq j\) is within the choice of ansatz (\ref{ansatz}) and so $T_{ij}$ is a diagonal matrix. In general, the spatial components of this diagonal matrix are not necessarily zero as we shall see for the BPS skyrmions with \(B>1\) in subsection \ref{subsection IIIa}.
%\ESkomen{\(T_{ij}=0\) untuk \(i\neq j\) karena \(\phi_i^a\phi^a_j=0\) untuk \(i\neq j\)}%\BGkomen{Apakah semua?}\ESkomen{untuk kasus statik hanya digunakan index \(00\) saja, pak.}
%
%%%%%%%
%%%%%NEW SECTION

\section{BPS Skyrme Submodels in the Spherically Symmetric Lagrangian Density}
\label{sec:BPS1}

In the spherically symmetric case, one can simplify the effective Lagrangian density \eqref{L_eff} to depend only on the radial coordinate $r$ by choosing \( f \left(  \theta  \right) =\tan  \theta  \) ,  \( g_{1} \left(  \varphi _{1} \right) = n_1 \varphi _{1} \) ,  \( g_{2} \left(  \varphi _{2} \right) = n_2 \varphi _{2} \), with $n_1, n_2\in\mathbb{Z}$.
It is necessary to impose a boundary condition \(  \xi \left( 0 \right) = m\pi \) with \(m\in\mathbb{Z}\) in order for all \(\phi\)'s to be unique at the origin. This boundary condition leads to the topological charge satisfying 
 \begin{equation}
     B=\frac{n_1 n_2}{16} \left( 9\cos  \left(  \xi_{\infty} \right) -\cos  \left( 3 \xi_{\infty} \right) \pm8 \right) ~ ,
 \end{equation} 
with  \(  \xi_{\infty}= \xi \left( r\rightarrow\infty \right)\), \(\left(+\right)\) sign is for odd \(m\), and \((-)\) for even \(m\).  Here we are only looking for the BPS skyrmions with $B\in\mathbb{Z}$ and we can choose to write $B=n_1n_2$. Therefore the boundary condition for \(  \xi \left( r\rightarrow\infty \right)  \) must satisfy the following equations
\begin{equation}  
9\cos  \left(  \xi_{\infty} \right) -\cos  \left( 3 \xi_{\infty} \right) \pm8 =16.
\end{equation}
If $m$ is odd then $\xi_{\infty}$ is equal to even multiples of \(\pi\), while even values of \(m\) gives us complex numbers. Here we will use the same boundary conditions as in~\cite{Skyrme:1962vh} for all submodels discussed in this paper, which are
 \begin{eqnarray}\label{boundcon}
   \xi \left( 0 \right) = \pi,\qquad \xi \left( r \rightarrow \infty \right) =0,
 \end{eqnarray} 
and searching for the BPS skyrmions with $B>0$ such that $n_1$ and $n_2$ must have the same positive or negative signs. Here, without loss of generality, we choose positive sign for both $n_1$ and $n_2$. Upon substituting \(f\), \(g_1\), and \(g_2\) mentioned above, the effective Lagrangian density (\ref{L_eff}) is simplified further to
\begin{eqnarray}  
L_{eff}&=& \left(  \xi'  \right) ^{2} \left[  \lambda _{1}+\left(1+n_1^2+n_2^2\right) \lambda _{2}\frac{\sin ^{2} \xi  }{r^{2}}+\left(n_1^2n_2^2+n_1^2+n_2^2\right) \lambda _{3}\frac{\sin ^{4} \xi  }{r^{4}}+ n_1^2n_2^2\lambda _{4}\frac{\sin ^{6} \xi  }{r^{6}} \right] \nonumber 
 \\ &&+ \left[ \lambda _{0}V+\left(1+n_1^2+n_2^2\right) \lambda _{1}\frac{\sin ^{2} \xi }{r^{2}}+\left(n_1^2n_2^2+n_1^2+n_2^2\right) \lambda _{2}\frac{\sin ^{4} \xi  }{r^{4}}+n_1^2n_2^2 \lambda_{3}\frac{\sin ^{6} \xi  }{r^{6}} \right].\label{eqn6}
\end{eqnarray}
Because the derivative terms in the effective Lagrangian density (\ref{eqn6}) contain only the first derivative of $\xi$, or \(  \xi'  \), the BPS-Lagrangian method in~\cite{Atmaja:2019gce} suggests that the corresponding \(L_{BPS}\) should take the following form
 \begin{equation}
     L_{BPS}=\frac{1}{r^{3}} \left( Q_{X}X+Q_{0} \right) ~ ,
 \end{equation} 
with  \( X\equiv \xi' \). In the BPS limit \( L_{eff}-L_{BPS}=0 \), 
\begin{eqnarray}
      \left[  \lambda _{1}+\left(1+n_1^2+n_2^2\right) \lambda _{2}\frac{\sin ^{2} \xi  }{r^{2}}+\left(n_1^2n_2^2+n_1^2+n_2^2\right) \lambda _{3}\frac{\sin ^{4} \xi  }{r^{4}}+ n_1^2n_2^2\lambda _{4}\frac{\sin ^{6} \xi  }{r^{6}} \right] X^{2}-\frac{1}{r^{3}}Q_{X}X \nonumber
\\
  + \left[ \lambda _{0}V+\left(1+n_1^2+n_2^2\right) \lambda _{1}\frac{\sin ^{2} \xi }{r^{2}}+\left(n_1^2n_2^2+n_1^2+n_2^2\right) \lambda _{2}\frac{\sin ^{4} \xi  }{r^{4}}+n_1^2n_2^2 \lambda_{3}\frac{\sin ^{6} \xi  }{r^{6}} -\frac{Q_{0}}{r^{3}}\right] =0 ~ ,\nonumber\\
\end{eqnarray}
which is a quadratic equation in \( X \) and has two solutions
 \begin{equation}
     X_\pm=\frac{Q_{X}}{2r^{3} \left[  \lambda _{1}+\left(1+n_1^2+n_2^2\right) \lambda _{2}\frac{\sin ^{2} \xi  }{r^{2}}+\left(n_1^2n_2^2+n_1^2+n_2^2\right) \lambda _{3}\frac{\sin ^{4} \xi  }{r^{4}}+ n_1^2n_2^2\lambda _{4}\frac{\sin ^{6} \xi  }{r^{6}} \right] } \pm \sqrt{D} ~ .\label{eqn2}
 \end{equation} 
These two solutions coincide, which is a necessary condition for Bogomolny's equations to be valid, if $D=0$,
\begin{eqnarray}
\frac{Q_{X}^{2}}{4r^{6}}- \left[  \lambda _{1}+\left(1+n_1^2+n_2^2\right) \lambda _{2}\frac{\sin ^{2} \xi  }{r^{2}}+\left(n_1^2n_2^2+n_1^2+n_2^2\right) \lambda _{3}\frac{\sin ^{4} \xi  }{r^{4}}+ n_1^2n_2^2\lambda _{4}\frac{\sin ^{6} \xi  }{r^{6}} \right] \nonumber\\
\times\left[ \lambda _{0}V+\left(1+n_1^2+n_2^2\right) \lambda _{1}\frac{\sin ^{2} \xi }{r^{2}}+\left(n_1^2n_2^2+n_1^2+n_2^2\right) \lambda _{2}\frac{\sin ^{4} \xi  }{r^{4}}+n_1^2n_2^2 \lambda_{3}\frac{\sin ^{6} \xi  }{r^{6}} -\frac{Q_{0}}{r^{3}}\right] =0 ~ .\label{eqn3}\nonumber\\
\end{eqnarray}
Since equation (\ref{eqn3}) must be satisfied for any value of  \( r \) then we may consider it as a polynomial equation of (explicit) $r$ in which each of its ``coefficients'' are zero. Non-trivial solutions can only be achieved by setting $Q_0=0$. The remaining vanishing ``coefficients'' are
 \begin{eqnarray}
  \lambda _{3} \lambda _{4}&=&0 ~ ,\label{c1}\\
 -\lambda _{2} \lambda _{4}&=& \lambda _{3}^{2} ~ ,\label{c2}\\
  -\lambda _{1}^{2}&=&V \lambda _{0} \lambda _{2} ~ ,\label{c3}\\
  -\left(n_1^2n_2^2+n_1^2+n_2^2\right) V \lambda _{0} \lambda _{3}&=&\left(\left(n_1^2n_2^2+n_1^2+n_2^2\right)+\left(1+n_1^2+n_2^2\right)^2\right) \lambda _{1} \lambda _{2} ~ ,\label{c4}\\
   - n_1^2n_2^2\left(1+n_1^2+n_2^2\right) \lambda _{1} \lambda _{4}&=& \left(n_1^2n_2^2\left(1+n_1^2+n_2^2\right)+\left(n_1^2n_2^2+n_1^2+n_2^2\right)^2\right) \lambda _{2} \lambda _{3} ~ ,\label{c5}\\
   \left(1+n_1^2+n_2^2\right)\left(n_1^2n_2^2+n_1^2+n_2^2\right) \lambda _{2}^{2}&+&\left(\left(1+n_1^2+n_2^2\right)\left(n_1^2n_2^2+n_1^2+n_2^2\right)+ n_1^2n_2^2\right) \lambda _{1} \lambda _{3}\nonumber\\+ n_1^2n_2^2 V \lambda _{0} \lambda _{4}  &=&\frac{Q_{X}^{2}}{4\sin ^{6} \xi \left( r \right)} ~ .\label{c6}
 \end{eqnarray}
There are two non-trivial solutions that simultaneously solve these equations. The first one is
\begin{equation}
 {Q^2_X\over 4}= \left(1+n_1^2+n_2^2\right)\left(n_1^2n_2^2+n_1^2+n_2^2\right)\lambda^2_2\sin^6\xi,
\end{equation}
with $\lambda_0=\lambda_1=\lambda_3=\lambda_4=0$, that, after substituting them into equation (\ref{eqn2}), gives us the Bogomolny equation
\begin{equation}
    \xi' = \pm \gamma(n_1,n_2) \frac{ \sin  \xi  }{r} ~ ,\label{eqn7}
\end{equation} 
where \(\gamma(n_1,n_2)\) is defined as
\begin{equation}\label{eq:gamma12}
    \gamma(n_1,n_2) \equiv \sqrt{\frac{n_1^2n_2^2+n_1^2+n_2^2}{1+n_1^2+n_2^2}} ~ .
\end{equation}
The second one is
\begin{equation}
 {Q^2_X\over 4}=n_1^2n_2^2V\lambda_0\lambda_4\sin^6\xi ~ ,
\end{equation}
with $\lambda_1=\lambda_2=\lambda_3=0$, whose corresponding Bogomolny equation is given by
\begin{equation}
     \xi' = \pm ~\sqrt[]{\frac{ \lambda _{0}}{ \lambda _{4}}}\frac{r^{3}}{n_1n_2\sin ^{3} \xi }\sqrt[]{V(\xi)} ~ .\label{eqn8}
\end{equation} 
Both cases show the dependence of the resulting Bogomolny's equation on the topological charge \(B=n_1n_2\). Let us discuss the features and solutions of those Bogomolny's equations in the next subsections.

\subsection{BPS Skyrme submodel with $\lambda_0=\lambda_1=\lambda_3=\lambda_4=0$}
\label{subsection IIIa}

Solutions to equation \eqref{eqn7} are
\begin{equation}\label{solquart}
    \xi=2\tan^{-1}\left[\left(\frac{r}{r_0}\right)^{\pm\gamma}\right] ~ ,
\end{equation}
with \(r_0\) is a real integration constant. Here we only pick the \((-)\) signed solutions that satisfy the boundary conditions \eqref{boundcon}, while the \((+)\) signed solutions are related to the anti-BPS skyrmions, with $B<0$. These solutions, with \(n_1=n_2=1\) or a unit topological charge \(B=1\), coincide with the solutions studied in \cite{Brihaye:2017wqa}. The solutions are smooth functions up to the second-order derivative in the whole domain, thus everywhere regular. The total static energy density of a BPS skyrmion, with topological charge $B$, in this model is given by 
\begin{equation}\label{energydensityquart}
    \varepsilon_B=-T^0_{0}=32\lambda_2\left(B^2+n^2+ 2B\right)\frac{r^{4(\gamma-1)}r_0^{4\gamma}}{\left(r^{2\gamma}+r_0^{2\gamma}\right)^4} ~ .
\end{equation}
with \(n \equiv n_2-n_1\). One may ask if these BPS Skyrmions saturate the total energy bound. We can show that they do saturate the energy bound at the effective level of total energy. For that purpose we write explicitly the Bogomolny equation \eqref{eqn7} as follows
\begin{equation}
  \sqrt{1+n^2+2B}\xi'\mp\sqrt{B^2+n^2+2B}\frac{\sin\xi}{r}=0.
\end{equation}
In the static case, the total energy is proportional to the action, \(E=\frac{\lambda_2}{4}\int\sqrt{-g} ~ d^4x ~ F^4\). Substituting the ansatz \eqref{ansatz} into \eqref{SEMT L2}, the total energy becomes
\begin{eqnarray}
    E&=&{\lambda_2\over 4}\int\sqrt{-g}d^4x \left[(1+n^2+2B)\xi'^2+(B^2+n^2+2B)\frac{\sin^2\xi}{r^2}\right]\frac{\sin^2\xi}{r^2}\nonumber\\
    &=& {\lambda_2\over 4}\int\sqrt{-g}d^4x\left[\left(\sqrt{1+n^2+2B}\xi'\mp\sqrt{B^2+n^2+2B}\frac{\sin\xi}{r}\right)^2\right.\nonumber\\&&\left.\pm2\sqrt{(1+n^2+2B)(B^2+n^2+2B)}\frac{\sin\xi}{r}\xi'\right]\frac{\sin^2\xi}{r^2}\nonumber\\
    &\geq&\pm{\lambda_2\over 2}\int\sqrt{-g}d^4x\sqrt{(1+n^2+2B)(B^2+n^2+2B)}\frac{\sin^3\xi}{r^3}\xi',   
\end{eqnarray}
where the last line is the boundary term.
Therefore, the total energy of the BPS Skyrmions is
 \begin{equation}
  E_B=\pm\lambda_2\pi^2\int_{\xi(\infty)}^{\xi(0)}\sqrt{(1+n^2+2B)(B^2+n^2+2B)}\sin^3\xi~d\xi.
 \end{equation}
The \(T^1_{1}\) component of the stress-energy-momentum tensor vanishes while the remaining diagonal components have the form
\begin{eqnarray}
T^2_{2}&=&2\frac{\sin^4\xi}{r^4}\frac{(n^2+2B)(1-B^2)}{1+n^2+2B} ~ ,\\
T^3_{3}&=&\frac{\sin^4\xi}{r^4}\left[n\sqrt{n^2+4B}(1+\gamma^2)+\frac{(n^2+2B)(B^2-1)}{1+n^2+2B}\right] ~ ,\\
T^4_{4}&=&\frac{\sin^4\xi}{r^4}\left[-n\sqrt{n^2+4B}(1+\gamma^2)+\frac{(n^2+2B)(B^2-1)}{1+n^2+2B}\right] ~ .
\end{eqnarray}

For $B=1$, which also implies $n=0$, all the remaining diagonal components of the energy-momentum tensor vanish, while they are non-zero for $B > 1$ in general. The latter cases indicate that we might have BPS skyrmions with non-zero pressures. Using \eqref{energydensityquart},  the total static energy of the BPS skyrmion is
\begin{equation}\label{mult sol}
    E_B=\frac{16}{3}\lambda_2\pi^2\sqrt{\left(1+n^2+ 2B\right)\left(B^2+n^2+ 2B\right)}  ~ .
\end{equation}
This is in agreement with the result found in \cite{Brihaye:2017wqa}, that the bound energy is linear in \(B\), because \(\frac{\sqrt{(1+n^2+2B)(B^2+n^2+2B)}}{3}\geq B\) and the equality occurs only for \(B=1\).
We can see a possible repulsive interaction in this BPS submodel between $B$ unit-solitons in the multi-soliton solution, with $B>1$, because the difference between the energy of the multi-soliton solution (\ref{mult sol}) and \(B\) times the unit-soliton energy, which is $B E_1=16\lambda_2\pi^2 B$, are positive and monotonically increasing. For example the energy of \(B=2\) multi-soliton is \(E_2=16\sqrt{6}~\lambda_2 \pi^2\) which is higher than the total energy of the two unit-solitons, \(2 E_1= 32\lambda_2 \pi^2\). This repulsive effect implies the possibility of a soliton creation in the usual collision set-up, similar to the one demonstrated in \cite{Vachaspati:2011ad}.

\begin{figure*}[ht!]
            \includegraphics[width=.45\textwidth]{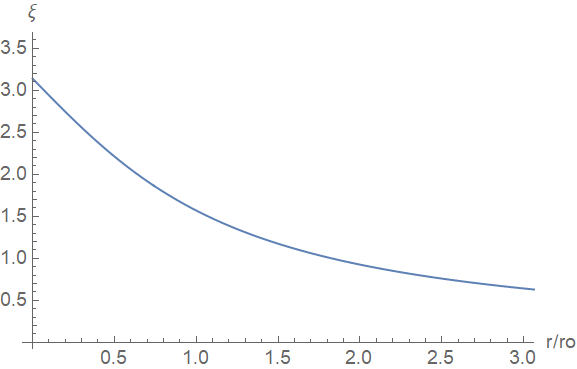}\hfill
            \includegraphics[width=.45\textwidth]{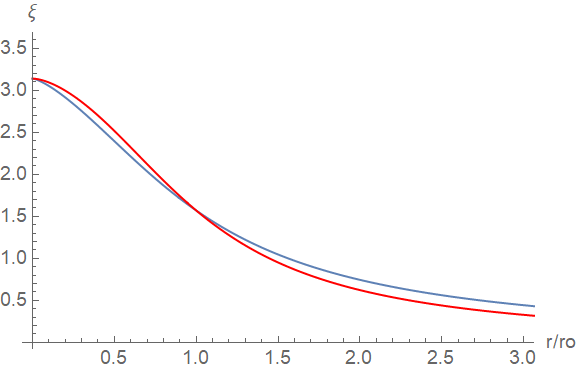}\hfill
                        \caption{Plots for \((-)\) sign solutions of equation \eqref{solquart} with topological charge \(B=1\) (left) and \(B=4\) (right). On the right plot, the blue curve is for solution with $E_4=16\sqrt{66}\lambda_2 \pi^2$ and the red curve is for solution with $E_4=32\sqrt{6}\lambda_2 \pi^2$.}
            \label{quart}
        \end{figure*}

\begin{figure*}[ht!]
            \includegraphics[width=.5\textwidth]{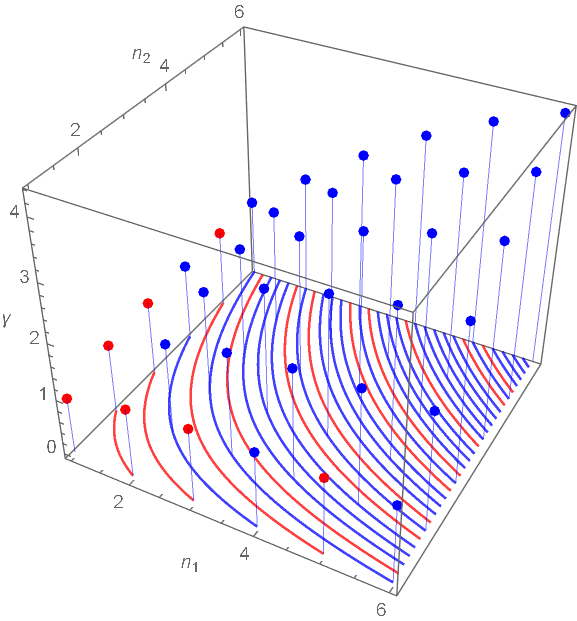}
            \includegraphics[width=.3\textwidth]{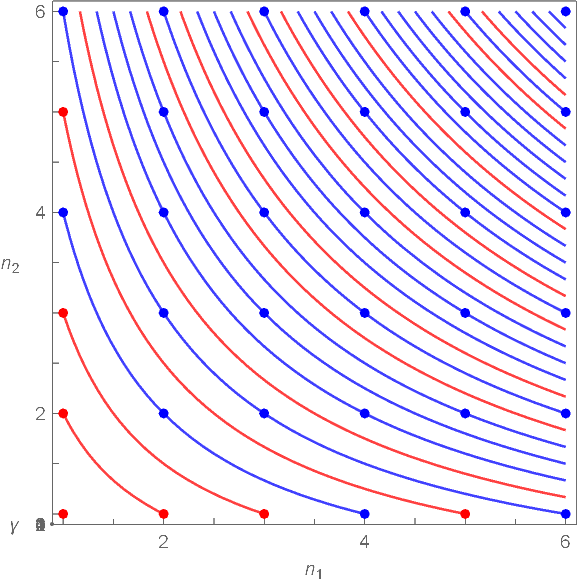}\hfill
            \caption{Plot for values of \(\gamma(n_1n_2)\). The red dots shows configurations in which the topological charges are prime numbers and the blue dots are for the topological degenerate configurations.}\label{gamma}
        \end{figure*}

Furthermore, for some fixed values of $B>1$, there are more than one BPS skyrmions in which their energies, determined by different values of $n_1$ and $n_2$, are not equal. As an example for \(B=4\)  equation \eqref{eqn7} has two different solutions and thus yields two unequal energies: $E_4=32\sqrt{6}\lambda_2 \pi^2$, with $n=2-2=0$, and $E_4=16\sqrt{66}\lambda_2 \pi^2$, with $n=1-4=-3$ and $n=4-1=3$. The graphical representation for this statement is given in figure \ref{quart}. We shall call these types of skyrmions ``topological degenerate'' BPS skyrmions which are defined as BPS skyrmions with the same topological charges but different energies. This feature arises from how many different ways we can assign integer values on \(n_1\) and \(n_2\) resulting in the same value of \(B\). Interestingly, there are also BPS skyrmions without these degenerate topological numbers in which their topological charges are prime numbers. 
\\   
The function \(\gamma(n_1,n_2)\) is plotted in Figure \ref{gamma} as a function of $n_1$ and $n_2$, with $n_1,n_2=1,\dots,10$. The curves with different colors connect all configurations of $n_1$ and $n_2$ with the same topological charge $B$.

\subsection{BPS Skyrme submodel with $\lambda_1=\lambda_2=\lambda_3=0$}\label{sec III-b}

In this submodel, solutions of the Bogomolny's equation (\ref{eqn8}) depend on the ratio of the coupling constants, \( \frac{ \lambda _{0}}{ \lambda _{4}} \), and also on the non-zero potential. In order to find the solutions of the Bogomolny's equation \eqref{eqn8}, we choose the potential to take the following form \(V(\xi)=\left(1-\cos \xi \right)^q \), with \(q\in\mathbb{R}^{+}\), such that the Bogomolny's equation \eqref{eqn8} can be rewritten as\footnote{Here again we pick the negative sign solutions that satisfy the boundary conditions \eqref{boundcon}.}
\begin{equation}\label{eqn12}
\sin^{3} \left(\frac{\xi}{2}\right)\cos ^{3} \left(\frac{\xi}{2}\right)\xi^{\prime}=- \sqrt{\frac{2^{2(q-3)}\lambda_{0}}{ \lambda_{4}}} \frac{r^{3} \sin^{q} \left(\frac{\xi}{2}\right)}{B}.
\end{equation}
The total static energy for this submodel, related to this form of potential, is given by
\begin{equation}\label{eqn26}
    E_B=128\pi^2\sqrt{2\lambda_0\lambda_4}\frac{B}{(q+4)(q+6)}.
\end{equation}

Now, let us introduce a dimensionless variable \( z= \left( \frac{2^{2(q-6)} \lambda _{0}}{\lambda _{4}} \right) ^{\frac{1}{2}}r^4 \) and a function \( u=\frac{ \xi}{2} \). Substituting them into equation \eqref{eqn12} yields
\begin{equation}\label{persxinonsferis1}
\frac{d u}{d z}=- \frac{\sin^{q-3} u}{B\cos ^{3} u}.
\end{equation}
A solution to this equation, that satisfies the boundary condition \eqref{boundcon} at the origin, $u={\pi\over 2}$, is given by
\begin{equation}\label{octicsolgeneral}
\frac{\sin ^{4-q} u}{4-q}-\frac{\sin ^{6-q} u}{6-q} =- \frac{z}{B}+ \frac{2}{\left(4-q\right)\left(6-q\right)}.
\end{equation}
Unfortunately, this solution does not satisfy the boundary condition \eqref{boundcon} at infinity, $u=0$, for $0<q<4$. For these values of $q$, the function $u$ is equal to zero at some finite point $z=z_{cut-off}$ and beyond this point $u$ becomes complex numbers. Therefore, for each value of $q$, we may define a cut-off point at $z=z_{cut-off}$ such that \(\xi(z_{cut-off})=0\), which is given by
\begin{equation}
    z_{cut-off}= \begin{cases} 
      \frac{2B}{(4-q)(6-q)} & 0<q<4 \\
      \infty & q \ge 4 
   \end{cases},
\end{equation}
Since the constant \(\xi=0\) is also a solution of the differential equation \eqref{eqn12}, we may then construct the compacton type solutions for $0<q<4$, while for $q\geq 4$ we have regular solutions. Unfortunately, it is difficult to write explicitly the function $u$, or $\xi$, from the solution \eqref{octicsolgeneral} in general and so we have to solve it numerically. Notice that  equation \eqref{octicsolgeneral} is singular for $q=4,6$ and so we can not use this equation as their solution. Below we will find their solutions starting by writing the explicit form of their Bogomolny's equations \eqref{persxinonsferis1}. 

Let us now consider the case of $q=4$ in which the potential is \( V(\xi)=\left(1-\cos \xi \right)^4 \). The Bogomolny's equation \eqref{persxinonsferis1} is explicitly written as
\begin{equation}\label{eqn14}
\frac{d u}{d z}=- \frac{ \sin u}{B\cos ^{3} u}
\end{equation}
The right hand side of equation \eqref{eqn14} is zero near infinity, \( z \rightarrow \infty \). This allows \(\xi(r)\) to converge at this limit, then satisfies the boundary condition near the infinity.
Solving equation \eqref{eqn14} and imposing the boundary condition at the origin, \(  u \left( 0 \right) = \pi/2   \), gives
\begin{equation}\label{eqn15}
\ln (\sin u)-\frac{1}{2} \sin ^{2} u+\frac{1}{2}=- \frac{z}{B}
\end{equation}
Unfortunately, equation \eqref{eqn15} is a transcendental equation for \(u(z)\), and so we have to use a numerical calculation to find its solution. 

\begin{figure}
   \centering
    \includegraphics{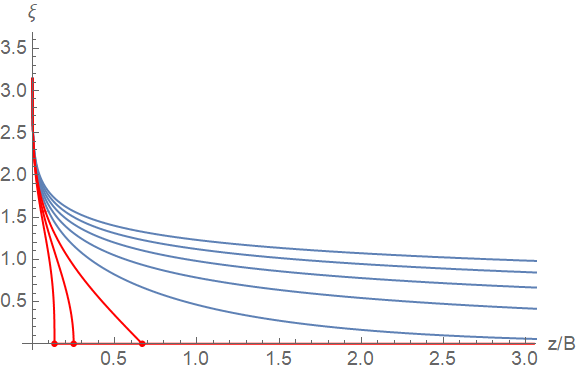}
    \caption{Numerical solutions of equation \eqref{persxinonsferis1} for \(q=1,\dots,8\) respectively from left to right curves. The red curves are the compacton solutions and the blue curves are the regular solutions.}
    \label{fig9}
\end{figure}

Another case is $q=6$ that has the potential \(V(\xi)=\left(1-\cos \xi  \right)^6 \). In this case the Bogomolny's equation \eqref{persxinonsferis1} becomes
\begin{equation}\label{eqn16}
\frac{d u}{d z}=- \frac{ \sin^{3} u}{B\cos ^{3} u}
\end{equation}
This equation has the same convergence feature as equation \eqref{eqn14} with a higher convergence rate at \(z\rightarrow\infty\). 
Integration on \eqref{eqn16} gives the solution we need. Again by imposing boundary condition at the origin, \(  u \left( 0 \right) = \pi/2   \), we get
\begin{equation}\label{eqn17}
-\ln (\sin u)-\frac{1}{2\sin ^{2} u} +\frac{1}{2}=- \frac{z}{B}
\end{equation}
It is easy to observe that equation \eqref{eqn17} is also transcendental in \(u(z)\) so we need to use numerical calculation to find its solution. All numerical calculations are given in Figure \ref{fig9}. We can see that an everywhere regular solution starts to emerge from \(q= 4\).
\\
%%%%%NEW SECTION

\section{BPS Skyrme Submodel in the Non-Spherically Symmetric Lagrangian Density} \label{sec:BPS2}

In this section, we consider the ansatz \eqref{ansatz} without fixing all the effective fields. It is easy to see that the effective Lagrangian \eqref{L_eff} also depends explicitly on the polar coordinate $\theta$ besides the radial coordinate $r$. In this case, we use the prescription in~\cite{Atmaja:2019gce} as such the general BPS Lagrangian density has the form
 \begin{equation} \label{LBPS}
 L_{BPS}=\frac{1}{r^{3}\sin  \theta \cos  \theta } \left(Q_{0}+\sum_{i=1}^{4}Q_{i}X_{i}
 +\sum_{i=1}^{3}\sum_{j=i+1}^{4}Q_{ij}X_{i}X_{j} \right) ~,
 \end{equation} 
where  \(X_{i}\) are functions of \(\xi'(r)\), \(f'(\theta)\), \(g_1'(\varphi_1)\) and \(g_2'(\varphi_2)\) to obtain a set of Bogomolny's equations.

The submodel discussed here is obtained by taking \(  \lambda _{1}= \lambda _{2}= \lambda _{3}=0 \) such that the Lagrangian \eqref{L_eff} simply becomes 
\begin{equation}
    L_{eff}=L_{0}+L_{4} = \lambda _{0}V \left(  \phi  \right) + \lambda _{4} \left[ \frac{1}{r^{6}\sin ^{2} \theta \cos ^{2} \theta }\frac{f^{2}}{ \left( 1+f^{2} \right) ^{4}}\sin ^{6} \xi  \left(  \xi' ~f' ~g_{1}' ~g_{2}'  \right) ^{2} \right] ~ .
\end{equation}  
Taking the \(L_{BPS}\) to be
\begin{equation} 
L_{BPS}=\frac{Q_{X}}{r^{3}\sin  \theta \cos  \theta } X ~ ,
\end{equation} 
with  \( X= \xi'f'g_{1}'g_{2}' \), the \( L_{eff}-L_{BPS}=0 \) becomes a quadratic equation of $X$, 
\begin{equation} 
 \left( \frac{ \lambda _{4}}{r^{6}\sin ^{2} \theta \cos ^{2} \theta }\frac{f^{2}}{ \left( 1+f^{2} \right) ^{4}}\sin ^{6} \xi \left( r \right)  \right) X^{2}- \left( \frac{Q_{X}}{r^{3}\sin  \theta \cos  \theta } \right) X+ \lambda _{0}V \left(  \phi  \right) =0 ~ ,
 \end{equation} 
which has solutions
\begin{equation} 
X_\pm= \left( \frac{Q_{X} \left( 1+f^{2} \right) ^{4}r^{3}\sin  \theta \cos  \theta }{2 \lambda _{4}f^{2}\sin ^{6} \xi \left( r \right) } \right)  \left( 1 \pm \sqrt[]{1-4 \left( \frac{ \lambda _{4}}{Q_{X}^{2}}\frac{f^{2}}{ \left( 1+f^{2} \right) ^{4}}\sin ^{6} \xi \left( r \right)  \right)  \lambda _{0}V} \right)  ~ .
\end{equation} 
The two solutions will coincide if
\begin{equation} 
4 \left( \frac{ \lambda _{4}}{Q_{X}^{2}}\frac{f^{2}}{ \left( 1+f^{2} \right) ^{4}}\sin ^{6} \xi \left( r \right)  \right)  \lambda _{0}V = 1 ~
\end{equation} 
which implies
\begin{equation}\label{Qx1}
 Q_{X}= \pm 2 \left( \frac{f}{ \left( 1+f^{2} \right) ^{2}}\sin ^{3} \xi \left( r \right)  \right) \sqrt[]{ \lambda _{0} \lambda _{4}V}  ~ 
 \end{equation}
and the corresponding Bogomolny's equation is
\begin{equation} \label{Bogomolnyeq}
 \xi^{'}f^{'}g_{1}^{'}g_{2}^{'} = \pm ~\sqrt[]{\frac{ \lambda _{0}}{ \lambda _{4}}}\frac{ \left( 1+f^{2} \right) ^{2}r^{3}\sin  \theta \cos  \theta }{f\sin ^{3} \xi  }\sqrt[]{V} ~ .
  \end{equation} 
The explicit form of \(L_{BPS}\) is given by
\begin{equation} L_{BPS}=\pm\frac{2 f\sin ^{3} \xi \left( r \right) \sqrt[]{ \lambda _{0} \lambda _{4}V}}{r^{3}\sin  \theta \cos  \theta \left( 1+f^{2} \right) ^{2}} \xi^{'} \left( r \right)f^{'}\left(  \theta  \right)g_{1}^{'}\left(  \varphi _{1} \right)g_{2}^{'}\left(  \varphi _{2} \right).  \end{equation} 
It is easy to show that no constraint equations are coming from the Euler-Lagrange equations of the above BPS Lagrangian density.

From the Bogomolny's equation \eqref{Bogomolnyeq}, we know that $g_{1}$  and  \( g_{2}\) have to be the linear functions $\varphi_1$ and $\varphi_2$ respectively. Therefore, similar to the spherically symmetric case, we take
\begin{equation}\label{sol varphi12}
g_{1}\left(  \varphi _{1} \right) =n_{1}~\varphi _{1},~g_{2}\left(  \varphi _{2} \right) =n_{2}~ \varphi _{2}, 
\end{equation} 
with $n_1,n_2>0$ are integer constants due to the uniqueness in coordinates $\varphi_{1,2}  \to \varphi_{1,2}  + 2\pi$. Using the Bogomolny's equation \eqref{Bogomolnyeq}, the total static energy density \(\varepsilon\) of the BPS skyrmion is given by
\\
\begin{equation}
    \varepsilon=-T_0^0=\sqrt[]{ \lambda _{0} \lambda _{4}V}\frac{2 f\sin ^{3} \xi    }{\left( 1+f^{2} \right) ^{2}r^{3}\sin  \theta \cos  \theta } \xi'f'g_{1}'g_{2}',
\end{equation}
and so the total static energy of the BPS skyrmion is
\begin{eqnarray}
E_B&=&\int_{domain-space} d^4x \sqrt{ \lambda _{0} \lambda _{4}V}\frac{2 f\sin ^{3} \xi    }{\left( 1+f^{2} \right) ^{2}} \xi' f' g_{1}'g_{2}',\nonumber\\
&=&2\sqrt{\lambda _{0} \lambda _{4}}\int_{target-space}\sqrt{V}\frac{ f\sin ^{3} \xi}{\left( 1+f^{2} \right) ^{2}} d\xi df dg_{1}dg_{2},\nonumber\\
&=&\frac{16\pi^2}{3}B\sqrt{\lambda _{0} \lambda _{4}}\left<\sqrt{V}\right>_{S^4}.\label{total energy}
\end{eqnarray}
All other diagonal components of the stress-energy-momentum tensor vanish for this BPS skyrmion. This formula for the total static energy is a generalization of the total static energy \eqref{eqn26}, that is for a particular form of the potential, in the spherically symmetric case. 

There are three types of potentials, using the ansatz \eqref{ansatz}, that are invariant under global $O(5)$ symmetry. The first one (Type-I) is the spherically symmetric potential with \(V \equiv V(\phi^5)=V(\cos\xi) \) in which the Bogomolny's equation \eqref{Bogomolnyeq} becomes
\begin{equation} \frac{f~f^{'}  }{ \left( 1+f^{2} \right) ^{2}\sin  \theta \cos  \theta }= \pm~ \sqrt[]{\frac{ \lambda _{0}}{ \lambda _{4}}}\frac{r^3 \sqrt{V(\cos\xi)}  }{ \xi^{'}  \sin ^3 \xi ~n_{1}n_{2}  }.\label{eqn20}
\end{equation} 
Using separation variables, we can solve this equation by first solving
\begin{equation}
    \xi' = -~\sqrt[]{\frac{ \lambda _{0}}{ \lambda _{4}}}\frac{r^{3}}{Bk\sin ^{3} \xi }\sqrt{V},
\end{equation}
with the topological charge is given by \(B=~n_1n_2\).
The solution to this equation has been discussed in section \ref{sec III-b} for a particular form of potential, which is equation \eqref{eqn12}.
\begin{equation} 
\frac{f~f^{'} }{ \left( 1+f^{2} \right) ^{2}\sin  \theta \cos  \theta } = \mp k ~ , \label{BE f type-I}
\end{equation} 
with \( k >0\) to satisfy boundary conditions for \(\xi\). The solution for this equation is
\begin{equation} 
f= \pm~ \sqrt[]{\frac{1}{c\mp k \cos^2\theta}-1 },\label{sol f}
 \end{equation} 
Putting this solution to \eqref{top charge} and imposing the choice for topological charge, \(B=n_1n_2\), yields a constraint \(k=\mp 1\) but we need to omit the negative one because \(k>0\). With this constraint, \(f\) on its boundaries takes value
\begin{eqnarray}
f(0)=\pm\sqrt{\frac{-c}{1+c}}\\
f(\frac{\pi}{2})=\pm\sqrt{\frac{1-c}{c}}
\end{eqnarray}
We need \(f\) to be real in this case, thus the only possibility is \(c=0\) which lead to
\begin{equation}
    f=\pm \tan\theta
\end{equation}
This result is just the spherically symmetric case we have in \ref{sec III-b}. The total static energy of this BPS skyrmion is given by equation \eqref{eqn26} for the case of \(V(\cos\xi)=\left(1-\cos\xi\right)^q\).

The Type-II potential is of the form $V\equiv V((\phi^1)^2+(\phi^2)^2)=V\left({f^2\over 1+f^2}\sin^2\xi\right)$. As an example we take $V={f^2\over 1+f^2}\sin^2\xi$ such that the Bogomolny's equation \eqref{Bogomolnyeq} becomes
\begin{equation}\label{eqn23}
    \frac{f^{'} }{ \left( 1+f^{2} \right) ^{3/2}\sin  \theta \cos  \theta }= \pm~ \sqrt[]{\frac{ \lambda _{0}}{ \lambda _{4}}}\frac{r^3}{ \xi^{'}  \sin ^2 \xi ~n_{1}n_{2}  } ~ .
\end{equation}
As previously, we solve it using separation variable as such the differential equation for \(f\) is given by
\begin{equation}\label{eqn24}
    \frac{f^{'} }{ \left( 1+f^{2} \right) ^{\frac{3}{2}}\sin  \theta \cos \theta }=\mp k ~ ,
\end{equation}
with $k$ is a positive separation constant. The corresponding solution is
\begin{equation}
    f=\pm\sqrt{\frac{(c_1\pm\frac{k}{2}\cos^2\theta)^2}{1-(c_1\pm\frac{k}{2}\cos^2\theta)^2}} ~ ,
\end{equation}
with \(c_1\) is a integration constant. Again, by imposing \(B=n_1n_2\) yields a relation \(c_1=\mp\frac{4+k^2}{4k}\). This relation can be used to evaluate \(f\) on its boundary
\begin{eqnarray}
f(0)=\pm\sqrt{\frac{\left(\frac{4-k^2}{4k}\right)^2}{1-\left(\frac{4-k^2}{4k}\right)^2}}=\pm\sqrt{\frac{\left(4-k^2\right)^2}{16k^2-\left(4-k^2\right)^2}}\\\label{eqn33}
f(\frac{\pi}{2})=\pm\sqrt{\frac{\left(\frac{4+k^2}{4k}\right)^2}{1-\left(\frac{4+k^2}{4k}\right)^2}}=\pm\sqrt{\frac{\left(4+k^2\right)^2}{-\left(4-k^2\right)^2}}
\end{eqnarray}
Because \(f\) is real valued, then from \eqref{eqn33} we can deduce that \(k=2\). Hence the solution for \(f\) is
\begin{equation}
    f=\pm\frac{\sin^2\theta}{\sqrt{1-\sin^4\theta}}
\end{equation}
differential equation for $\xi$ is now
\begin{equation}\label{eqn25}
    \sin^3\xi~\xi'=- \sqrt{\frac{\lambda_0}{\lambda_4}} \frac{r^3}{Bk}\sin\xi,
\end{equation}
Solving equation \eqref{eqn25} give us a compacton solution
\begin{equation}\label{eqn32}
    \xi-\frac{\sin(2\xi)}{2}=- \sqrt{\frac{\lambda_0}{\lambda_4}} \frac{r^4}{4B}+\pi,
\end{equation}
in which the cut-off is at $r_{cut-off}=\left(4\sqrt{\lambda_4\over\lambda_0}{B}\pi\right)^{1/4}$.
The total static energy for this BPS skyrmion is given by 
\begin{equation}\label{eqn35}
    E_B=\pi^3\sqrt{\lambda_0\lambda_4}B.
\end{equation}

\begin{figure*}[ht!]
            \includegraphics[width=.50\textwidth]{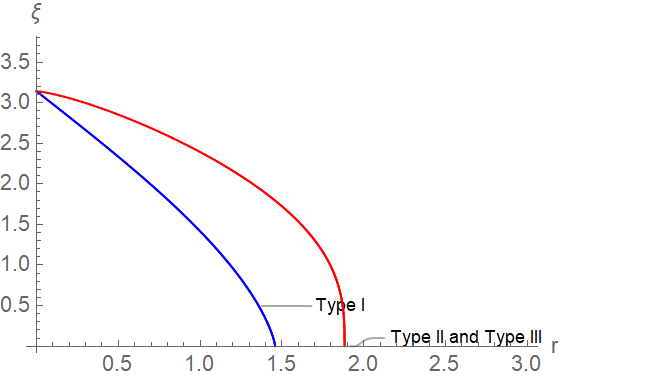}\hfill
            \includegraphics[width=.50\textwidth]{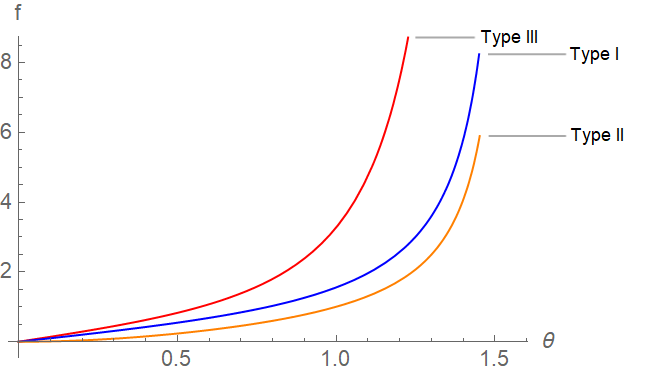}\hfill
                        \caption{Plots of  solutions for three types of potential with the same total energy of non-spherically symmetric submodel with only potential and octic term. Solution for \(\xi\) (left) is the same for the second and third type, but solution for \(f\) is different for every type of potential (right).}
            \label{non-spher}
        \end{figure*}

Another possible potential is of the form $V\equiv V((\phi^3)^2+(\phi^4)^2)=V\left({\sin^2\xi\over 1+f^2}\right)$ which is the Type-III potential. Again, as an example, we take $V={\sin^2\xi\over 1+f^2}$ such that the Bogomolny's equation \eqref{Bogomolnyeq} becomes
\begin{equation}\label{eqn31}
    \frac{ff^{'} }{ \left( 1+f^{2} \right) ^{3/2}\sin  \theta \cos  \theta }= -~ \sqrt[]{\frac{ \lambda _{0}}{ \lambda _{4}}}\frac{r^3}{ \xi^{'}  \sin ^2 \xi ~n_{1}n_{2}  } ~ .
\end{equation}
As previously, solving the first equation for the angular field \(f\),
\begin{equation}
    \frac{ff^{'} }{ \left( 1+f^{2} \right) ^{3/2}\sin  \theta \cos  \theta }=\mp k,\label{BE f type-III}
\end{equation}
with $k$ is a positive constant. The same steps from the previous cases can be applied to this case yields
\begin{equation}
    f=\pm \sqrt{\frac{1-(c_2\pm\frac{k}{2}\cos^2\theta)^2}{(c_2\pm\frac{k}{2}\cos^2\theta)^2}} ~ ,
\end{equation}
with \(c_2\) is an integration constant. Again, by imposing \(B=n_1n_2\) yields a relation \(c_2=\pm\frac{4-k^2}{4k}\). Hence the boundary condition for \(f\) is
\begin{eqnarray}
f(0)=\pm\sqrt{\frac{1-\left(\frac{4+k^2}{4k}\right)^2}{\left(\frac{4+k^2}{4k}\right)^2}}=\pm\sqrt{\frac{-\left(4-k^2\right)^2}{\left(4+k^2\right)^2}}\\\label{eqn34}
f(\frac{\pi}{2})=\pm\sqrt{\frac{1-\left(\frac{4-k^2}{4k}\right)^2}{\left(\frac{4-k^2}{4k}\right)^2}}=\pm\sqrt{\frac{16k^2-\left(4-k^2\right)^2}{\left(4-k^2\right)^2}}
\end{eqnarray}
From \eqref{eqn34} we can deduce that \(k=2\) which lead to a specific solution of \(f\), namely
\begin{equation}
    f=\pm\frac{\sqrt{1-\cos^4\theta}}{\cos^2\theta}
\end{equation}
\\
The differential equation for \(\xi\) for this case is the same as the one from the previous case which is given in \eqref{eqn25}, thus the solution is given by \eqref{eqn32}. This solution has total static energy given by \eqref{eqn35}. According to the prescription in~\cite{Atmaja:2019gce}, there could be other possible (genuine) BPS submodels of the effective Lagrangian density \eqref{L_eff}, which consist of four derivative terms. However, we found that those submodels do not have Bogomolny's equations as shown in the appendix. To sum up this section, the plot of \(\xi\) and positive signed \(f\) solutions for the same total energy of all three types of potentials is given in figure \ref{non-spher}.

\section{Conclusions}
\label{sec:concl}

We have considered some possible (generic) BPS submodels of the generalized five-dimensional Skyrme Model via the BPS Lagrangian method. The analysis is mainly focused on finding Bogomolny's Equations and their solutions, with topological charge $B$ is a positive integer, and also the total static energy of these BPS skyrmions under a particular ansatz~\eqref{ansatz}. The two main cases that have been considered, based on the effective Lagrangian density of the BPS submodels, are the spherically symmetric and the non-spherically symmetric cases. Employing the BPS Lagrangian method, we followed the prescription in~\cite{Atmaja:2019gce} for the BPS Lagrangian density.
In the spherically symmetric case, where only the effective field $\xi$ in the ansatz~\eqref{ansatz} is undetermined, we found there are only two possible BPS submodels which are the submodel with \(\lambda_0=\lambda_1=\lambda_3=\lambda_4=0\), or namely the Skyrme term, and the submodel with \(\lambda_1=\lambda_2=\lambda_3=0\), the potential term and a term that is proportional to the square of the topological current.

The first BPS submodel with only the Skyrme term has been studied in \cite{Brihaye:2017wqa} for a unit BPS skyrmion solution in which $B=1$. Here we extended the study to the BPS skyrmion solutions with $B>1$. We found two distinct features compared to the corresponding four-dimensional BPS Skyrme model discussed in~\cite{Harland:2013rxa}. The first one is that the BPS skyrmions exist only if there is no potential which is in contradiction with the four-dimensional case in which the BPS skyrmions may exist if the potential takes some specific forms~\cite{Harland:2013rxa,Atmaja:2019gce}. The second one is the existence of ``topological degenerate'' BPS skyrmions defined as BPS skyrmions with the same topological charge, but different total static energies; except for the unit BPS skyrmion, $B=1$. Interestingly there are also BPS skyrmions with the same topological charge and energies only if the topological charge is the prime number. Furthermore, the BPS skyrmions, with $B>1$, have non-zero pressures at least in two angular directions while the unit BPS skyrmion has no pressures. The BPS skyrmions, with $B>1$, have total static energy that is higher than the sum of $B$ times unit BPS skyrmion which indicates that these BPS skyrmions are repulsive and unstable. It is interesting to investigate the dynamics of soliton scattering in this submodel because the repulsive effect implies the possibility of soliton creation in the usual collision setup, similar to the one demonstrated in \cite{Vachaspati:2011ad}. We may need to add gravitational force, by incorporating gravity into the BPS submodel, to overcome the pressures so that these BPS skyrmions, with $B>1$, could be stable.  This will be studied further in the future article.

The second BPS submodel depends on the specific form of the potential term. In this article, we considered the potential term of the form \(V=(1-\cos\xi)^q\), with \(q\in\mathbb{R}^{+}\). We found the BPS skyrmions are compacton solutions for $0<q<4$ and are regular solutions for $q\geq 4$. The compacton solutions have a finite cut-off radius proportional to \(B^{1/4}\). The total static energy of BPS skyrmions in this submodel is proportional to the topological charge $B$ similar to the BPS-Skyrme model in four dimensions \cite{Adam:2010zz}. This may be because both models consist of the potential term and the term that is proportional to the square of the topological current, $B^\mu B_\mu$.

In the non-spherically symmetric case, in which all the effective fields in the ansatz~\eqref{ansatz} are undetermined, we found the only possible BPS submodel is similar to the second BPS submodel in the spherically symmetric case. In this case, the potential term could depend on the polar coordinate $\theta$, and hence the effective field $f$ could have general forms which depend on the symmetry of the potential term. This BPS submodel gives a generalization of the corresponding spherically symmetric one in which the general form of the total static energy~\eqref{total energy} is similar to the one found in \cite{Adam:2010zz}. The results found in this more general scheme cover the preceding spherically symmetric case.

\acknowledgments
BEG  would like to acknowledge Eugene Radu for email correspondence related to the paper \cite{Brihaye:2017wqa}. We also thank M. Satriawan for careful reading the paper and  improving English grammar in the paper. The work in this paper is supported by Kemendikbudristek-ITB WCR 2021. BEG would also thank ITB for financial support through  Riset ITB 2021.

\appendix
%%%

\section{Full Field Equations of Motion}
This section is devoted to showing that all the results in this paper indeed satisfy the corresponding full-field equations of motion of the generalized five-dimensional Skyrme model derived from the Lagrangian \eqref{eq:L}. Let us first write down the action related to \eqref{eq:L} as
\begin{equation}
    \mathcal{S} = \int\sqrt{-g} ~ d^5x \left[\lambda _{0}V \left(  \phi  \right) + \lambda _{1}F^{2}+\frac{ \lambda _{2}}{4}F^{4}+\frac{ \lambda _{3}}{36}F^{6}+\frac{ \lambda _{4}}{576}F^{8}\right] ~ ,
\end{equation}
with \(F^{2n}=\phi^{a_1}_{[i_1}\dots\phi^{a_n}_{i_n]}\phi^{a_1}_{[j_1}\dots\phi^{a_n}_{j_n]}g^{i_1j_1}\dots g^{i_nj_n}\) satisfy \(O(5)\) model constraint \(\phi^a\phi^a=1\). The full field equations of motion for this model are given by
\begin{eqnarray}\label{15}
    &&\left(\delta^{gf}-\phi^g\phi^f\right)\left[2\lambda_1\nabla_n\phi^{f,n}+2\lambda_2\nabla_n\left(\phi^a_i\phi^{a,i}\phi^{f,n}-\phi^a_i\phi^{f,i}\phi^{a,n}\right)\right.\nonumber\\&&\left.\lambda_3\nabla_n\left((\phi^a_i\phi^{a,i})^2\phi^{f,n}-\phi^a_i\phi^{a,j}\phi^b_j\phi^{b,i}\phi^{f,n}-2\phi^a_i\phi^{a,i}\phi^b_j\phi^{f,j}\phi^{b,n}+2\phi^f_i\phi^{a,i}\phi^a_j\phi^{b,j}\phi^{b,n}\right)\right.\nonumber\\&&\left.\frac{\lambda_4}{72}\nabla_n\left(B_m\varepsilon^{ijknm}\varepsilon^{abcfe}\phi_i^a\phi_j^b\phi_k^c\phi^e\right)\right]=\lambda_0\frac{\partial V(\phi)}{\partial\phi^f}~.
\end{eqnarray}

\subsection{BPS Skyrme submodel with \(\lambda_0=\lambda_1=\lambda_3=\lambda_4=0\)}
For the genuine submodel discussed in \cite{Brihaye:2017wqa} with \(\lambda_0=\lambda_1=\lambda_3=\lambda_4=0\) the Lagrangian becomes
\begin{eqnarray}
    \mathcal{L} = \frac{\lambda_2}{4}F^4=\frac{\lambda_2}{4}\phi^{ab}_{i_1i_2}\phi^{ab}_{j_1j_2}g^{i_1j_1}g^{i_2j_2}  ~ ,
  \end{eqnarray}
which leads to the following equations of motion
\begin{eqnarray}\label{2}
    (\delta^{da}-\phi^d\phi^a)\left(\phi^{a,k}\partial_k(\phi^b_i\phi^{b,i})+\phi^b_i\phi^{b,i}\nabla_k\phi^{a,k}-\phi^{b,k}\partial_k(\phi^a_i\phi^{b,i})-\phi^a_i\phi^{b,i}\nabla_k\phi^{b,k}\right)=0 ~ .
\end{eqnarray}
Using the spherically symmetric ansatz fields \eqref{ansatz} and the bipolar spherical metric \eqref{eq:metricanz}, we can simplify some of the terms in \eqref{2}
\begin{eqnarray}\label{3}
    \phi^b_i\phi^{b,i}&=& \xi'^{2}+(1+n_1^2+n^2_2)\frac{\sin ^{2} \xi }{r^{2}} ~ ,\\ \label{4}
    \phi^a\nabla_k\phi^{a,k}&=& -\left(\xi'^{2}+(1+n_1^2+n^2_2)\frac{\sin ^{2} \xi }{r^{2}}\right) ~ ,\\ \label{5}
    \phi^a\phi^{b,k}\partial_k(\phi^a_i\phi^{b,i})&=&-\left(\xi'^{4}+(1+n_1^4+n^4_2)\frac{\sin ^{4} \xi }{r^{2}}\right) ~ ,\\ \label{6}
    \phi^a\phi^a_i&=&0~\text{for}~i\in\{0,1,2,3,4\} ~ ,\\ \label{7}
    \phi^a_i\phi^a_j&=&0~\text{for}~i\neq j ~ ,
\end{eqnarray}
such that each of equations in (\ref{2}) can be simplified to
\begin{eqnarray}\label{8}
     && \frac{\sin\xi}{r^4}\left(\left(n_1^2+n_2^2+1\right)r^2\xi'^2\sin(2\xi)-\left(n_1^2+n_2^2+1\right)r\sin^2\xi\left(\xi'+3r\xi'^2\cot\xi+r\xi''\right)\right.\nonumber\\&&\left.+\left(\left(n_2^2+1\right)n_1^2+n_2^2\right)\sin(2\xi)\sin^2\xi\right) = 0 ~ ,
\end{eqnarray}
multiplied by a non-zero multiplication factor.
Suppose we have the solution of \(\xi\) which satisfies the first order equation \(\xi'=\gamma \frac{\sin\xi}{r}\) with \(\gamma\) is an arbitrary nonzero real constant. Substituting this first-order equation, equation \eqref{8} becomes
\begin{equation}\label{9}
    -\frac{32\left(\gamma^2+\left(\gamma^2-1\right)n_2^2+n_1^2\left(\gamma^2-n_2^2-1\right)\right)r^{4\gamma-4}\left(r^{2\gamma}-1\right)}{\left(r^{2\gamma}+1\right)^5} = 0 ~ ,
\end{equation}
that must be valid for all \(r\) and so we have
\begin{equation}
    \gamma^2(n_1^2+n_2^2+1)-(n_1^2+n_2^2+n_1^2n_2^2)=0 ~ .
\end{equation}
This leads to \eqref{eq:gamma12} which implies that the Bogomolny equation (\ref{eqn7}) satisfies the full field equations of motion of the BPS submodel.

\subsection{BPS Skyrme submodel with $\lambda_1=\lambda_2=\lambda_3=0$}
The next genuine submodel with octic and potential terms has \(\lambda_1=\lambda_2=\lambda_3=0\) whose Lagrangian is given by
\begin{eqnarray}
    \mathcal{L}&=&\frac{\lambda_4}{576}F^8+\lambda_0V\nonumber\\
    &=&\frac{\lambda_4}{576}\phi^{abcd}_{i_1i_2i_3i_4}\phi^{abcd}_{j_1j_2j_3j_4}g^{i_1j_1}g^{i_2j_2}g^{i_3j_3}g^{i_4j_4}+\lambda_0 V ~ .
\end{eqnarray}
To simplify the calculation we use the fact that the octic term is the norm of the topological current (before the normalization)
\begin{equation}
    B^m=\varepsilon^{ijklm}\varepsilon^{abcde}\phi^a_{i}\phi^b_{j}\phi^c_{k}\phi^d_{l}\phi^e ~ ,
\end{equation}
satisfy \(F^8=B_m B^m\). For the non-spherically symmetric cases in the submodel with the octic term and the potential term, we have the equation of motion as
\begin{eqnarray}\label{14}
\left(\delta^{gf}-\phi^g\phi^f\right)\left[\frac{2\lambda_4}{576}\nabla_n\left(B_m\frac{\partial B^m}{\partial \phi^f_n}\right)-\lambda_0\frac{\partial V(\phi)}{\partial\phi^f}\right] = 0 ~ ,
\end{eqnarray}
which can be further simplified to
\begin{equation}
    \left(\delta^{gf}-\phi^g\phi^f\right)\left[\frac{8\lambda_4}{576}\nabla_n\left(B_m\varepsilon^{ijknm}\varepsilon^{abcfe}\phi_i^a\phi_j^b\phi_k^c\phi^e\right)-\lambda_0\frac{\partial V(\phi)}{\partial\phi^f}\right]=0 ~ . \label{eq:EOMlg}
\end{equation}
%%%%
Next, we consider three types of potential, that is, \(V(\phi^5)\), \(V(\phi^1,\phi^2)\), and \(V(\phi^3,\phi^4)\).  
%%%%%
\subsubsection{Type I potential \(V(\phi^5)\)}
%%%
First, we consider the case in which the potential is chosen to be \(V(\phi^5)=V(\xi(r))\). Upon substituting the ansatz (\ref{ansatz}) and the Bogomolny equation (\ref{BE f type-I}) for $f$ into the full field equations \eqref{eq:EOMlg}, we obtain that each of equation \eqref{eq:EOMlg} can be simplified as
\begin{equation}\label{11}
    \frac{2 \lambda _4 k^2 n_1^2 n_2^2 \sin ^7\xi \left(3 \xi ' \left(r \xi ' \cot \xi
   -1\right)+r \xi ''\right)}{r^7}+\lambda _0 \frac{\partial V}{\partial\phi^5} \sin ^2\xi = 0 ~ .
\end{equation}
multiplied by a nonzero multiplication factor. Substituting further the Bogomolny equation (\ref{eqn12}) for $\xi$ with general potential $V(\xi(r))$, that is, $\xi'=-\sqrt{\lambda_0\over\lambda_4}{r^3\sqrt{V}\over n_1n_2k\sin^3\xi}$, into equation \eqref{11} gives us a trivial result. Therefore we may conclude that the Bogomolny equations \eqref{BE f type-I} and \eqref{eqn12} satisfy the full field equations \eqref{eq:EOMlg}.

\subsubsection{Type II Potential \(V(\phi^1,\phi^2)\) and Type III Potential \(V(\phi^3,\phi^4)\) }

The second type of potential is chosen explicitly to be \(V(\phi^1,\phi^2)=V((\phi^1)^2+(\phi^2)^2)={f^2\over 1+f^2}\sin^2\xi\). Upon substituting the ansatz \eqref{ansatz} and the Bogomolny equation \eqref{eqn24} for $f$ into equations \eqref{eq:EOMlg}, we get a simplified equation 
\begin{equation}\label{12}
    \lambda _0 r^7 \cos \xi -\lambda _4 k^2 n_1^2 n_2^2 \sin ^4\xi 
   \left(-3 \xi ' \sin \xi +3 r \left(\xi '\right)^2 \cos \xi +r \xi '' \sin \xi \right)=0~
\end{equation}
for each of equations \eqref{eq:EOMlg} with a non-zero multiplication factor. The next step is to substitute the Bogomolny equation \eqref{eqn25} for $\xi$ into equation \eqref{12} which give us another trivial result. Therefore again the Bogomolny equations \eqref{eqn24} and \eqref{eqn25} satisfy the full field equations \eqref{eq:EOMlg}.

\subsubsection{Type III potential, \(V(\phi^3,\phi^4)\) }

The third type of potential is chosen explicitly to be \(V(\phi^3,\phi^4)=V((\phi^3)^2+(\phi^4)^2)={1\over 1+f^2}\sin^2\xi\). Again substituting the ansatz \eqref{ansatz} and the Bogomolny equation \eqref{BE f type-III} for $f$ into equations \eqref{eq:EOMlg} implies the same simplified equation \eqref{12} which is also turn out to be trivial after substituting further the Bogomolny equation \eqref{eqn25} for $\xi$. Therefore also the Bogomolny equations \eqref{BE f type-III} and \eqref{eqn25} satisfy the the full field equations \eqref{eq:EOMlg}.

%\newpage

\section{Submodels With Four Derivative Terms}
The possible BPS submodels with four number of derivative terms can be obtained by setting \( \lambda _{1}= \lambda _{2}= \lambda _{4}=0 \) or \(  \lambda _{2}= \lambda _{3}= \lambda _{4}=0 \) in the effective Lagrangian density~\eqref{L_eff}. Both of them are possible because we can express all derivative terms in four \(X_i\)'s. On the other hand, the BPS Lagrangian method does not apply to submodel with only quartic terms which have \(\lambda _{0}=\lambda _{1}= \lambda _{3}= \lambda _{4}=0\) in this choice of ansatz because we cannot transform the derivative terms into four \(X_i\)'s.
%%%
\subsection{Submodel with \(  \lambda _{1}= \lambda _{2}= \lambda _{4}=0 \)}

Taking \(  \lambda _{1}= \lambda _{2}= \lambda _{4}=0 \), the effective Lagrangian density \eqref{L_eff} becomes
\begin{eqnarray}
 L_{eff}&=&L_{0}+L_{3}\nonumber\\ 
&=& \lambda _{0}V \left(  \phi  \right) + \lambda _{3} \left[ \frac{1}{r^{4}\sin ^{2} \theta }\frac{  f^2  }{ \left( 1+f^{2} \right) ^{3}}\sin ^{4} \xi  \left(  \xi' f' g_{1}' \right) ^{2}+\frac{1}{r^{4}\cos ^{2} \theta }\frac{1}{ \left( 1+f^{2}  \right) ^{3}}\sin ^{4} \xi  \left(  \xi'f'g_{2}' \right) ^{2}\right.\nonumber \\
&&\left. +\frac{1}{r^{6}\sin ^{2} \theta \cos ^{2} \theta }\frac{  f^2 }{ \left( 1+f^{2}  \right) ^{4}}\sin ^{6} \xi \left( f'g_{1}'g_{2}' \right) ^{2}+\frac{1}{r^{4}\sin ^{2} \theta \cos ^{2} \theta }\frac{  f^2 }{ \left( 1+f^{2}  \right) ^{2}}\sin ^{4} \xi  \left(  \xi'g_{1}'g_{2}' \right) ^{2} \right] .\nonumber\\
\end{eqnarray}
A suitable form of the BPS Lagrangian is given by \eqref{LBPS} with  \( X_1= \xi'g_{1}'g_{2}' \) ,  \( X_2=f'g_{1}'g_{2}'\) , \( X_3=\xi'f'g_{2}'\) , and  \(X_4=\xi'f'g_{1}'\). We set \( L_{eff}-L_{BPS}=0 \) and then solve it as a quadratic equation of $X_1,X_2,X_3,$ and $X_4$ subsequently. Demanding each $X_i$'s solutions to be equal we then get an algebraic equation that does not contain any derivative of the effective fields and may explicitly depend on the coordinates, namely $r$ and $\theta$. Rewriting this equation as a polynomial equation of (explicit) $r$, we then set each of its ``coefficients'' to be zero such that we obtain the following equations
\begin{eqnarray}
   &&-Q_0 Q_{14}^2 Q_{23}^2 f^8-Q_0 Q_{13}^2 Q_{24}^2 f^8+2 Q_1 Q_3 Q_{13} Q_{24}^2
   f^8-Q_0 Q_{12}^2 Q_{34}^2 f^8\nonumber\\&&+2 Q_3^2 Q_{12} Q_{14} Q_{24} f^8-2 Q_1 Q_3 Q_{14} Q_{23} Q_{24} f^8+2 Q_0 Q_{13}
   Q_{14} Q_{23} Q_{24} f^8\nonumber\\&&+2 Q_0 Q_{12} Q_{14} Q_{23} Q_{34} f^8-2 Q_1 Q_3 Q_{12} Q_{24} Q_{34} f^8+2 Q_0 Q_{12}
   Q_{13} Q_{24} Q_{34} f^8\nonumber\\&&+2 Q_1^2 Q_{23} Q_{24} Q_{34} f^8-4 Q_0 Q_{14}^2 Q_{23}^2 f^6-4 Q_0 Q_{13}^2 Q_{24}^2 f^6+8
   Q_1 Q_3 Q_{13} Q_{24}^2 f^6-4 Q_0 Q_{12}^2 Q_{34}^2 f^6\nonumber\\&&+8 Q_3^2 Q_{12} Q_{14} Q_{24} f^6-8 Q_1 Q_3 Q_{14} Q_{23}
   Q_{24} f^6+8 Q_0 Q_{13} Q_{14} Q_{23} Q_{24} f^6\nonumber\\&&+8 Q_0 Q_{12} Q_{14} Q_{23} Q_{34} f^6-8 Q_1 Q_3 Q_{12} Q_{24}
   Q_{34} f^6\nonumber\\&&+8 Q_0 Q_{12} Q_{13} Q_{24} Q_{34} f^6+8 Q_1^2 Q_{23} Q_{24} Q_{34} f^6-6 Q_0 Q_{14}^2 Q_{23}^2 f^4-6 Q_0
   Q_{13}^2 Q_{24}^2 f^4\nonumber\\&&+12 Q_1 Q_3 Q_{13} Q_{24}^2 f^4-6 Q_0 Q_{12}^2 Q_{34}^2 f^4+12 Q_3^2 Q_{12} Q_{14} Q_{24}
   f^4-12 Q_1 Q_3 Q_{14} Q_{23} Q_{24} f^4\nonumber\\&&+12 Q_0 Q_{13} Q_{14} Q_{23} Q_{24} f^4+12 Q_0 Q_{12} Q_{14} Q_{23} Q_{34}
   f^4-12 Q_1 Q_3 Q_{12} Q_{24} Q_{34} f^4\nonumber\\&&+12 Q_0 Q_{12} Q_{13} Q_{24} Q_{34} f^4+12 Q_1^2 Q_{23} Q_{24} Q_{34} f^4-4
   Q_0 Q_{14}^2 Q_{23}^2 f^2\nonumber\\&&-4 Q_0 Q_{13}^2 Q_{24}^2 f^2+8 Q_1 Q_3 Q_{13} Q_{24}^2 f^2-4 Q_0 Q_{12}^2 Q_{34}^2 f^2+8
   Q_3^2 Q_{12} Q_{14} Q_{24} f^2\nonumber\\&&-8 Q_1 Q_3 Q_{14} Q_{23} Q_{24} f^2+8 Q_0 Q_{13} Q_{14} Q_{23} Q_{24} f^2+8 Q_0 Q_{12}
   Q_{14} Q_{23} Q_{34} f^2\nonumber\\&&-8 Q_1 Q_3 Q_{12} Q_{24} Q_{34} f^2+8 Q_0 Q_{12} Q_{13} Q_{24} Q_{34} f^2+8 Q_1^2 Q_{23}
   Q_{24} Q_{34} f^2\nonumber\\&&-2 V \sin ^6(\xi ) Q_{13} Q_{14} Q_{34} \lambda _0 \lambda _3 f^2-Q_0 Q_{14}^2 Q_{23}^2-Q_0
   Q_{13}^2 Q_{24}^2\nonumber\\&&+2 Q_1 Q_3 Q_{13} Q_{24}^2-Q_0 Q_{12}^2 Q_{34}^2+2 \left(f^2+1\right)^4 Q_4^2 Q_{12} Q_{13}
   Q_{23}+2 Q_3^2 Q_{12} Q_{14} Q_{24}\nonumber\\&&-2 Q_1 Q_3 Q_{14} Q_{23} Q_{24}+2 Q_0 Q_{13} Q_{14} Q_{23} Q_{24}+2
   \left(f^2+1\right)^4 Q_2^2 Q_{13} Q_{14} Q_{34}\nonumber\\&&+2 Q_0 Q_{12} Q_{14} Q_{23} Q_{34}-2 Q_1 Q_3 Q_{12} Q_{24} Q_{34}+2
   Q_0 Q_{12} Q_{13} Q_{24} Q_{34}\nonumber\\&&+2 Q_1^2 Q_{23} Q_{24} Q_{34}+2 \left(f^2+1\right)^4 Q_4 \left(\left(Q_2 Q_{13}-Q_1
   Q_{23}\right) \right.\nonumber\\&&\left.\left(\right.Q_{13} Q_{24}-Q_{14} Q_{23}\right)-Q_{12} \left(Q_2 Q_{13}+Q_1 Q_{23}\right) Q_{34}+Q_3 Q_{12}
   \left(\right.-Q_{14} Q_{23}-Q_{13} Q_{24}\nonumber\\&&+Q_{12} Q_{34}\left.\right)\left.\right)+2 \left(f^2+1\right)^4 Q_2 \left(Q_3 Q_{14}
   \left(Q_{14} Q_{23}-Q_{13} Q_{24}\right.\right.\nonumber\\&&\left.\left.-Q_{12} Q_{34}\right)+Q_1 Q_{34} \left(-Q_{14} Q_{23}-Q_{13} Q_{24}+Q_{12}  Q_{34}\right)\right) = 0 ~ ,
   \end{eqnarray} \pagebreak
\begin{eqnarray}
   && 4 \csc ^2(2 \theta ) Q_2^2 Q_{34}^2 f^{12}+8
   \csc ^2(2 \theta ) Q_0 Q_{23} Q_{24} Q_{34} f^{12}+\csc ^2(\theta ) Q_2^2 Q_{13}^2 f^{10}\nonumber\\&&+\csc ^2(\theta ) Q_1^2
   Q_{23}^2 f^{10}+20 \csc ^2(2 \theta ) Q_2^2 Q_{34}^2 f^{10}-2 \csc ^2(\theta ) Q_1 Q_2 Q_{13} Q_{23} f^{10}\nonumber\\&&+2 \csc
   ^2(\theta ) Q_0 Q_{12} Q_{13} Q_{23} f^{10}+40 \csc ^2(2 \theta ) Q_0 Q_{23} Q_{24} Q_{34} f^{10}\nonumber\\&&+4 \csc ^2(\theta )
   Q_2^2 Q_{13}^2 f^8+\sec ^2(\theta ) Q_2^2 Q_{14}^2 f^8+4 \csc ^2(\theta ) Q_1^2 Q_{23}^2 f^8\nonumber\\&&+\sec ^2(\theta ) Q_1^2
   Q_{24}^2 f^8+40 \csc ^2(2 \theta ) Q_2^2 Q_{34}^2 f^8-8 \csc ^2(\theta ) Q_1 Q_2 Q_{13} Q_{23} f^8\nonumber\\&&+8 \csc ^2(\theta
   ) Q_0 Q_{12} Q_{13} Q_{23} f^8-2 \sec ^2(\theta ) Q_1 Q_2 Q_{14} Q_{24} f^8\nonumber\\&&+2 \sec ^2(\theta ) Q_0 Q_{12} Q_{14}
   Q_{24} f^8+80 \csc ^2(2 \theta ) Q_0 Q_{23} Q_{24} Q_{34} f^8+6 \csc ^2(\theta ) Q_2^2 Q_{13}^2 f^6\nonumber\\&&+4 \sec ^2(\theta
   ) Q_2^2 Q_{14}^2 f^6+6 \csc ^2(\theta ) Q_1^2 Q_{23}^2 f^6+4 \sec ^2(\theta ) Q_1^2 Q_{24}^2 f^6\nonumber\\&&+40 \csc ^2(2 \theta
   ) Q_2^2 Q_{34}^2 f^6-12 \csc ^2(\theta ) Q_1 Q_2 Q_{13} Q_{23} f^6\nonumber\\&&+12 \csc ^2(\theta ) Q_0 Q_{12} Q_{13} Q_{23}
   f^6-8 \sec ^2(\theta ) Q_1 Q_2 Q_{14} Q_{24} f^6\nonumber\\&&+8 \sec ^2(\theta ) Q_0 Q_{12} Q_{14} Q_{24} f^6+80 \csc ^2(2 \theta
   ) Q_0 Q_{23} Q_{24} Q_{34} f^6\nonumber\\&&+4 \csc ^2(\theta ) Q_2^2 Q_{13}^2 f^4+6 \sec ^2(\theta ) Q_2^2 Q_{14}^2 f^4+4 \csc
   ^2(\theta ) Q_1^2 Q_{23}^2 f^4\nonumber\\&&+6 \sec ^2(\theta ) Q_1^2 Q_{24}^2 f^4+20 \csc ^2(2 \theta ) Q_2^2 Q_{34}^2 f^4-8 \csc
   ^2(\theta ) Q_1 Q_2 Q_{13} Q_{23} f^4\nonumber\\&&+8 \csc ^2(\theta ) Q_0 Q_{12} Q_{13} Q_{23} f^4-12 \sec ^2(\theta ) Q_1 Q_2
   Q_{14} Q_{24} f^4\nonumber\\&&+12 \sec ^2(\theta ) Q_0 Q_{12} Q_{14} Q_{24} f^4+40 \csc ^2(2 \theta ) Q_0 Q_{23} Q_{24} Q_{34}
   f^4+\csc ^2(\theta ) Q_2^2 Q_{13}^2 f^2\nonumber\\&&+4 \sec ^2(\theta ) Q_2^2 Q_{14}^2 f^2+\csc ^2(\theta ) Q_1^2 Q_{23}^2 f^2+4
   \sec ^2(\theta ) Q_1^2 Q_{24}^2 f^2\nonumber\\&&+4 \csc ^2(2 \theta ) Q_2^2 Q_{34}^2 f^2-2 \csc ^2(\theta ) Q_1 Q_2 Q_{13} Q_{23}
   f^2+2 \csc ^2(\theta ) Q_0 Q_{12} Q_{13} Q_{23} f^2\nonumber\\&&-8 \sec ^2(\theta ) Q_1 Q_2 Q_{14} Q_{24} f^2+8 \sec ^2(\theta )
   Q_0 Q_{12} Q_{14} Q_{24} f^2\nonumber\\&&+\left(f^2+1\right)^4 \csc ^2(\theta ) Q_3^2 \left(Q_{12}^2+\left(f^2+1\right) \sec
   ^2(\theta ) Q_{24}^2\right) f^2\nonumber\\&&+8 \csc ^2(2 \theta ) Q_0 Q_{23} Q_{24} Q_{34} f^2-2 \left(f^2+1\right)^4 \csc
   ^2(\theta ) Q_3 \nonumber\\&&\left(\left(f^2+1\right) Q_{24} \left(Q_4 Q_{23}+Q_2 Q_{34}\right) \sec ^2(\theta )+Q_{12} \left(Q_2
   Q_{13}+Q_1 Q_{23}\right)\right) \nonumber\\&&f^2-V \sin ^6(\xi ) \left(\right.\left(\csc ^2(\theta ) Q_{13}^2+4 \left(f^2+1\right) \csc
   ^2(2 \theta ) Q_{34}^2\right) f^2\nonumber\\&&+\sec ^2(\theta ) Q_{14}^2\left.\right) \lambda _0 \lambda _3 f^2+\sec ^2(\theta ) Q_2^2
   Q_{14}^2+\sec ^2(\theta ) Q_1^2 Q_{24}^2\nonumber\\&&+\left(f^2+1\right)^4 Q_4^2 \left(\sec ^2(\theta ) Q_{12}^2+4 f^2
   \left(f^2+1\right) \csc ^2(2 \theta ) Q_{23}^2\right)\nonumber\\&&-2 \sec ^2(\theta ) Q_1 Q_2 Q_{14} Q_{24}+2 \sec ^2(\theta )
   Q_0 Q_{12} Q_{14} Q_{24}\nonumber\\&&+2 \left(f^2+1\right)^4 Q_4 \left(-4 f^2 \left(f^2+1\right) Q_2 Q_{23} Q_{34} \csc ^2(2
   \theta )-\sec ^2(\theta ) Q_{12} \left(Q_2 Q_{14}+Q_1 Q_{24}\right)\right) = 0 ~ , \nonumber\\
    \end{eqnarray}
    \begin{eqnarray}
  &&f^2 \left(f^2+1\right) Q_{23} \left(2 Q_2
   Q_3-Q_0 Q_{23}\right) \csc ^2(\theta )\nonumber\\&&+\left(f^2+1\right) Q_{24} \left(2 Q_2 Q_4-Q_0 Q_{24}\right) \sec ^2(\theta
   )-Q_0 Q_{12}^2+2 Q_1 Q_2 Q_{12}\nonumber\\
   &&-\left(f^2+1\right)^4 f^2 \lambda _3^2 Q_2^2 \sin
   ^6(\xi )+f^{16} Q_1^2 Q_{34}^2+2 f^{16} Q_0 Q_{13} Q_{14} Q_{34}\nonumber\\&&+8 f^{14} Q_1^2 Q_{34}^2+16 f^{14} Q_0 Q_{13} Q_{14}
   Q_{34}+28 f^{12} Q_1^2 Q_{34}^2+56 f^{12} Q_0 Q_{13} Q_{14} Q_{34}\nonumber\\&&+56 f^{10} Q_1^2 Q_{34}^2+112 f^{10} Q_0 Q_{13}
   Q_{14} Q_{34}+70 f^8 Q_1^2 Q_{34}^2+140 f^8 Q_0 Q_{13} Q_{14} Q_{34}\nonumber\\&&+56 f^6 Q_1^2 Q_{34}^2+112 f^6 Q_0 Q_{13} Q_{14}
   Q_{34}+28 f^4 Q_1^2 Q_{34}^2+56 f^4 Q_0 Q_{13} Q_{14} Q_{34}\nonumber\\&&+8 f^2 Q_1^2 Q_{34}^2+16 f^2 Q_0 Q_{13} Q_{14}
   Q_{34}+\left(f^2+1\right)^8 Q_4^2 Q_{13}^2+\left(f^2+1\right)^8 Q_3^2 Q_{14}^2\nonumber\\&&-2 \left(f^2+1\right)^8 Q_1 Q_3 Q_{14}
   Q_{34}-2 \left(f^2+1\right)^8 Q_4 Q_{13} \left(Q_3 Q_{14}+Q_1 Q_{34}\right)\nonumber\\&&+f^4 \lambda _0 \lambda _3^3 V \sin
   ^{12}(\xi )+Q_1^2 Q_{34}^2+2 Q_0 Q_{13} Q_{14} Q_{34} = 0 ~,\\
  && Q_{14}^2 Q_{23}^2-2 Q_{14} \left(Q_{13} Q_{24}+Q_{12} Q_{34}\right) Q_{23}+\left(Q_{13}
   Q_{24}-Q_{12} Q_{34}\right)^2 = 0 ~ ,\\
   &&2 Q_{12} \left(f^2 Q_{13} Q_{23}
   \cos ^2(\theta )+Q_{14} Q_{24} \sin ^2(\theta )\right)+2 \left(f^2+1\right) f^2 Q_{23} Q_{24}
   Q_{34} = 0 ~ , \label{eqn29} \\
   &&Q_{12}^2+(1+f^2)(f^2\frac{Q_{23}^2}{\sin^2\theta}+\frac{Q_{24}^2}{\cos^2\theta}) = 0 ~ ,\\
  &&4 Q_1 \left(f^2 Q_3 Q_{13} \cos
   ^2(\theta )+Q_4 Q_{14} \sin ^2(\theta )\right)-2 Q_0 \left(f^2 \left(\left(f^2+1\right) Q_{34}^2+Q_{13}^2 \cos
   ^2(\theta )\right)\right.\nonumber\\&&\left.+Q_{14}^2 \sin ^2(\theta )\right)+4 \left(f^2+1\right) f^2 Q_3 Q_4 Q_{34}  = 0 ~ ,\label{eqn28}\\
  &&Q_1^2+(1+f^2)(f^2\frac{Q_3^2}{\sin^2\theta}+\frac{Q_4^2}{\cos^2\theta}) = 0 ~ ,\\
  &&Q_0 = 0 ~ ,
\end{eqnarray}  
To solve those equations, we may further rewrite them as polynomial equations of \(\sin\theta\) and set all their ``coefficients'' to be zero. Then, we have 
\begin{equation}\label{eqn30sol}
    Q_1= Q_2 = Q_3 = Q_4 =  Q_{12}=Q_{23} =Q_{24}= Q_{34}=Q_{13}=Q_{14}=0 ~ ,
\end{equation}
which eventually leads to trivial solutions. Therefore this submodel does not have Bogomolny's equation and so there is no BPS skyrmion.

\subsection{Submodel with \(  \lambda _{2}= \lambda _{3}= \lambda _{4}=0 \)}

In this submodel, the effective Lagrangian density \eqref{L_eff} becomes
\begin{eqnarray}
L_{eff}&=&L_{0}+L_{1}\nonumber\\ 
&=& \lambda _{0}V \left(  \phi  \right) + \lambda _{1} \left[  \left(  \xi'  \right) ^{2}+\frac{\sin ^{2} \xi }{r^{2}} \left( \frac{f' }{1+f^{2} } \right) ^{2}+\frac{\sin ^{2} \xi  }{r^{2}\sin ^{2} \theta ~}\frac{ \left( f  g_{1}' \right) ^{2}}{1+f^{2} }+\frac{\sin ^{2} \xi  }{r^{2}\cos ^{2} \theta ~}\frac{ \left( ~g_{2}' ~ \right) ^{2}}{1+f^{2}  } \right]\nonumber\\
\end{eqnarray}
Again, a suitable form of the BPS Lagrangian is given by \eqref{LBPS} with  \( X_1= \xi'\) ,  \( X_2=f'\) , \( X_3=g_{1}'\) and  \(X_4=g_{2}'\). Setting \( L_{eff}-L_{BPS}=0 \) and solving it as a polynomial equation of (explicit) $r$, we obtain one of the ``coefficient'' equations is 
%one of the resulting equation is
\begin{equation}
\lambda_1^4\lambda_0Vf^2\sin^6\xi=0     ~ .
\end{equation}
%%%%%%
Here the only possible solution is \(\lambda_0=0\), otherwise the submodel is non-dynamical, such that the remaining equations, which are written as polynomials in \(\sin\theta\), become
\begin{eqnarray}
(1+f^2)f^2Q_2^2+\sin^2\theta Q_3^2+f^2\cos^2\theta Q_4^2 &=& 0 ~, \\
(1+f^2)Q_2(\sec^2\theta Q_3Q_{23}+f^2\csc^2\theta Q_4Q_{24})+Q_3Q_4Q_{34} &=& 0 ~ ,\\
\sin^2\theta Q_1Q_{13}+f^2((1+f^2)Q_2Q_{12}+\cos^2\theta Q_4Q_{14}) &=& 0 ~ ,\\
(1+f^2)\sec^2\theta (Q_2Q_{13}-Q_1Q_{23})^2 \nonumber\\
+f^2(1+f^2)\csc^2\theta (Q_4^2Q_{12}^2+(Q_2Q_{14}-Q_1Q_{24})^2-2Q_4Q_{12}(Q_2Q_{14}+Q_1Q_{24}))\nonumber\\
-2Q_3((1+f^2)\sec^2\theta Q_{12}(Q_2Q_{13}+Q_1Q_{23})+Q_{14}(Q_4Q_{13}+Q_1Q_{34}))\nonumber\\
+Q_3^2((1+f^2)\sec^2\theta Q_{12}^2+Q_{14}^2)+(Q_4Q_{13}-Q_1Q_34)^2 &=& 0 ~ .
\end{eqnarray}
Taking
\begin{equation}
    Q_2=Q_3=Q_4=Q_{23}=Q_{24}=Q_{34}=0\label{eqn27}
\end{equation}
we obtain that it could be  either \(Q_1=0\) or \(\lambda_1=0\). This fact leads to a conclusion that this submodel cannot satisfy the BPS limit which shows the non-existence of Bogomolny's equation.

\medskip
$\bibliographystyle{utphys}$
\bibliography{skyrme5d}

\end{document}